\documentstyle[12pt,a4]{article}

\newcommand{\text}{\rm}

\begin{document}

\title{\textbf{Zero Curvature Formalism of \ the 4D Yang-Mills Theory in Superspace}
}
\author{\textbf{L.C.Q.Vilar, S.P. Sorella} \\
%EndAName
UERJ, Universidade do Estado do Rio de Janeiro\\
Departamento de F\'\i sica Te\'orica\\
Instituto de F\'{\i}sica\\
Rua S\~ao Francisco Xavier, 524\\
20550-013, Maracan\~a, Rio de Janeiro, Brazil\vspace{2mm}\\
and\vspace{2mm} \and \textbf{C.A.G.Sasaki } \\
%EndAName
C.B.P.F \\
Centro Brasileiro de Pesquisas F\'\i sicas \\
Rua Xavier Sigaud 150, 22290-180 Urca \\
Rio de Janeiro, Brazil \and \textbf{Ref. UERJ/DFT-03/98 }\vspace{2mm}\newline
\and \textbf{PACS: 11.10.Gh}}
\maketitle

\begin{abstract}
The supersymmetric descent equations in $N=1$ superspace are discussed by
means of the introduction of two operators $\zeta ^\alpha $, $\overline{%
\zeta }^{\dot \alpha }$ which allow to decompose the supersymmetric
covariant derivatives $D^\alpha $, $\overline{D}^{\dot \alpha }$ as BRS
commutators.

\setcounter{page}{0}\thispagestyle{empty}
\end{abstract}

\vfill\newpage\ \makeatother
\def\thesection {\Roman{section}}
\renewcommand{\theequation}{\thesection.\arabic{equation}}

\section{\ Introduction\-}

It is well known nowadays that the problem of finding the anomalies and the
invariant counterterms which arise in the renormalization of local field
theories can be handled in a purely algebraic way by means of the BRS
technique\footnote{%
For a recent account on the so called \textit{Algebraic Renormalization }see 
\cite{book}.}. This amounts to look at the nontrivial solution of the
integrated consistency condition 
\begin{equation}
s\;\int \omega _D^g\;=\;0\;,  \label{brst-cons-cond}
\end{equation}
where $s$ is the BRS operator and $g$ and $D$ denote respectively the ghost
number and the space-time dimension. Condition (\ref{brst-cons-cond}) when
translated at the nonintegrated level yields a system of equations usually
called descent equations (see \cite{book} and refs. therein) 
\begin{equation}
\begin{array}{l}
\,s\;\omega _D^g\;+\;d\;\omega _{D-1}^{g+1}\;=\;0\;, \\ 
\\ 
s\;\omega _{D-1}^{g+1}\;+\;d\;\omega _{D-2}^{g+2}\;=\;0\;, \\ 
\;\;\;\;\;\;\;\;\;\;..... \\ 
\;\;\;\;\;\;\;\;\;\;..... \\ 
s\;\omega _1^{g+D-1}\;+\;d\;\omega _0^{g+D}=\;0\;, \\ 
\\ 
s\;\omega _0^{g+D}=\;0\;,
\end{array}
\label{non-susy-desc-equat}
\end{equation}
where $d=dx^\mu \;\partial _\mu \;$is the exterior space-time derivative and 
$\omega _i^{g+D-i}\;(0\leq i\leq D)$ are local polynomials in the fields of
ghost number $(g+D-i)$ and form degree $i$. The cases $g=0,1$ correspond
respectively to the invariant counterterms and to the anomalies. The
operators $s$, $d$ obey the algebraic relations 
\begin{equation}
s^2\,=\,d^2\,=\,s\,d+d\,s\,=0\ .  \label{s-d-algebra}
\end{equation}
\ \ \ \ \ \ The problem of solving the descent equations (\ref
{non-susy-desc-equat}) is a problem of cohomology of $s$ modulo $d$\cite
{Brandt,Barnich}, the corresponding cohomology classes being given by
solutions of (\ref{non-susy-desc-equat}) which are not of the type 
\begin{eqnarray*}
\omega _m^{g+D-m} &=&s\,\hat{\omega}_m^{g+D-m-1}+d\,\hat{\omega}%
_{m-1}^{g+D-m}\ ,\qquad 1\leq m\leq D\ , \\
\omega _0^{g+D} &=&s\,\hat{\omega}_0^{g+D-1}\ ,
\end{eqnarray*}
with $\hat{\omega}$'s local polynomials. Notice also that at the
nonintegrated level one loses the property of making integration by parts.
This implies that the fields and their derivatives have to be considered as
independent variables.

Of course, the knowledge of the most general nontrivial solution of the
descent equations (\ref{non-susy-desc-equat}) yields the integrated
cohomology classes of the BRS operator. Indeed, once the full system (\ref
{non-susy-desc-equat}) has been solved, integration on space-time gives the
general solution of the consistency condition (\ref{brst-cons-cond}).

Recently, a new method of obtaining nontrivial solutions of the tower (\ref
{non-susy-desc-equat}) has been proposed\cite{Sorella} and successfully
applied to a large number of field models such as Yang-Mills theories\cite
{Sorella,Tataru}, gravity\cite{Gravity}, topological field theories\cite
{Topol1,Topol2,Topol3}, string\cite{Corda} and superstring\cite{Supercorda}
theories as well as $W_3$-algebras\cite{W3}. The method relies on the
introduction of an operator $\delta $ which allows to decompose the exterior
derivative as a BRS commutator, 
\begin{equation}
d\;=\;-\;\left[ s\;,\;\delta \right] \ .  \label{delt-def}
\end{equation}
It is easily proven in fact that repeated applications of the operator $%
\delta $ on the cocycle $\omega _0^{g+D}$ which solves the last of the
equations (\ref{non-susy-desc-equat}) will provide an explicit nontrivial
solution for the higher cocycles $\omega _i^{g+D-i}$.

One has to note that solving the last equation of the tower (\ref
{non-susy-desc-equat}) is a problem of local cohomology instead of a modulo-$%
d\;$one. The former can be systematically analysed by using several methods
as, for instance, the spectral sequences technique\cite{Dixon}. It is also
worth to mention that in the case of the Yang-Mills type gauge theories the
solutions of the descent equations (\ref{non-susy-desc-equat}) obtained via
the decomposition (\ref{delt-def}) have been proven to be equivalent to
those provided by the so called \textit{Russian Formula}\cite{Russ1,Russ2}.

Another important geometrical aspect related to the existence of the
operator $\delta \;$is the possibility of encoding all the relevant
informations concerning the BRS transformations of the fields and the
solutions of the system (\ref{non-susy-desc-equat}) into a unique equation
which takes the form of a generalized zero curvature condition\cite
{zero-curv}, \textit{i.e.} 
\begin{equation}
\widetilde{\mathcal{F}}\;=\;\widetilde{d}\;\widetilde{\mathcal{A}}\;-%
\widetilde{\mathcal{A}}^2\;=\;0\;.  \label{nonsusy-zero-curv-cond}
\end{equation}
The operator $\tilde{d}\;$and the generalized gauge connection $\widetilde{%
\mathcal{A}}\;$in eq.(\ref{nonsusy-zero-curv-cond}) turn out to be
respectively the $\delta $-transform of the BRS operator $s\;$and of the
ghost field $c\;$corresponding to the Maurer-Cartan form of the underlying
gauge algebra 
\begin{eqnarray*}
\widetilde{d}\; &=&\;e^\delta se^{-\delta }\;,\;\;\;\;\;\;\;\;\;\widetilde{d}%
^2\;=\;0\;, \\
\widetilde{\mathcal{A}}\; &=&\;e^\delta c\;.
\end{eqnarray*}
As discussed in detail in refs.\cite{zero-curv} the zero curvature
formulation allows to obtain straightforwardly the cohomology classes of the
operator ${\tilde{d}}$.${\;}$The latters are deeply related to the solutions
of the descent equations (\ref{non-susy-desc-equat}).

The BRS algebraic procedure can be easily adapted to include the case of the
renormalizable N=1 superspace supersymmetric gauge theories in four
space-time dimensions, for which a set of superspace descent equations have
been established\cite{pig1,pig2,pig3}. The solution of these equations as
much as in the nonsupersymmetric case yields directly all the manifestly
supersymmetric gauge anomalies as well as the manifestly supersymmetric BRS
invariant counterterms. One has to remark however that in the supersymmetric
case both the derivation and the construction of a solution of the
superspace version of the descent equations are more involved than the
nonsupersymmetric case, due to the algebra of the spinorial covariant
derivatives $D_\alpha \;$and $\overline{D}_{\dot{\alpha}}$ and to the
(anti)chirality constraints of some of the superfieds characterizing the
theory.

In order to have an idea of the differences between the superspace and the
ordinary case, let us briefly consider the integrated superspace N=1 BRS
consistency condition corresponding to the supersymmetric chiral U(1)
Yang-Mills axial anomaly\cite{pig3}

\begin{equation}
s\int d^4x\;d^2\overline{\theta }\;K^0{{\;}}=\;0\;,  \label{susyWZ}
\end{equation}
with $K^0$ a local power series in the gauge vector superfield with ghost
number zero and dimension two. It can be proven \cite{pig2,pig3} that
condition (\ref{susyWZ}) implies that the BRS variation of the integrand, 
\textit{i.e }$sK^0$, is a total derivative in superspace 
\begin{equation}
s\;K^0\;=\;\;\overline{D}_{\dot{\alpha}}\overline{K}^{\;1\dot{\alpha}}\;,\;
\label{local-susyWZ}
\end{equation}
$\;$with $\overline{K}^{\;1\dot{\alpha}}$ local power series with ghost
number one\footnote{%
The absence of the term $D^\alpha K_\alpha ^1$ in eq.(\ref{local-susyWZ}) is
actually due to the chirality nature of the consistency condition (\ref
{susyWZ}).}. Acting now on both sides of eq.(\ref{local-susyWZ}) with the
nilpotent BRS operator $s$ we get 
\[
\;\;\overline{D}_{\dot{\alpha}}\;s\overline{K}^{\;1\dot{\alpha}}=\;0\;. 
\]
This equation admits a superspace solution (see Sect.V and App.A for the
details) which, as in the standard nonsuperspace case (\ref
{non-susy-desc-equat}), entails a set of new conditions which together with
the equation (\ref{local-susyWZ}) gives the whole set of the superspace
descent equations for the U(1) axial anomaly\cite{pig3}, namely 
\begin{equation}
\begin{array}{lll}
s\;K^0 & = & \overline{D}_{\dot{\alpha}}\overline{K}_{\;}^{1\;\dot{\alpha}%
}\;, \\ 
&  &  \\ 
s\;\overline{K}_{\;\dot{\alpha}}^1\; & = & \left( 2D^\alpha \overline{D}_{%
\dot{\alpha}}\;+\;\overline{D}_{\dot{\alpha}}D^\alpha \right) \;K_{\;\alpha
}^{2\;}\;, \\ 
&  &  \\ 
s\;K^{2\;\alpha } & = & D^\alpha \;K^3\;, \\ 
&  &  \\ 
s\;K^3\; & = & \;0\;,
\end{array}
\label{desc-chiral0}
\end{equation}
with $K_{\;\alpha }^{2\;}$ and $K^3$ local power series of ghost number two
and three.

From now on, the operator $s$ in eqs.(\ref{desc-chiral0}), whose explicit
form will be given later in Sect.II (see eqs.(\ref{brs})), refers to the
BRS operator acting on the space of superfields of $N=1$ Yang-Mills theories
in superspace. Therefore, its integrated cohomology in the sectors of ghost
number 1 and 0 will yield the possible gauge anomalies and gauge invariant
counterterms which, being given in terms of superfields, are manifestly
supersymmetric. In particular, the eqs.(\ref{desc-chiral0}) are the superspace 
analogous of the Wess-Zumino consistency condition (\ref{non-susy-desc-equat}%
) for the U(1) axial anomaly.

For the sake of clarity it is worth emphasizing here that the existence of
anomalies for the supersymmetry itself has been deeply investigated by \cite
{D, B}, who showed in fact that the $N=1$ supersymmetric algebra with the
usual content of superfields (\textit{i.e.}, scalar, chiral and vector
superfields) does not allow for anomalies of supersymmetry.

The aim of this paper is to extend the previous works \cite{zero-curv} to
the case of the N=1 four dimensional supersymmetric Yang-Mills theory,
yielding thus a simple way of solving the superspace descent equations. This
means that we will introduce two operators $\zeta _\alpha \;$and $\overline{%
\zeta }_{\dot{\alpha}}\;$which in analogy with the case of the operator $%
\delta \;$of eq.(\ref{delt-def}) allow to decompose the supersymmetric
covariant derivatives $D_\alpha \;$and $\overline{D}_{\dot{\alpha}}\;$as BRS
commutators, according to 
\begin{equation}
\left[ \zeta _\alpha \;,\;s\right] \;=\;D_\alpha \,,\qquad \left[ \overline{%
\zeta }_{\dot{\alpha}}\;,\;s\right] \;=\;\overline{D}_{\dot{\alpha}}\ ,
\label{ss-dec}
\end{equation}
with 
\begin{equation}
D_\alpha \,\overline{D}_{\dot{\alpha}}\,+\,\overline{D}_{\dot{\alpha}%
}\,D_\alpha \,=\,2i\sigma _{\alpha \dot{\alpha}}^\mu \;\partial _\mu \;\,,
\label{susy-algebra1}
\end{equation}
$\sigma _{\alpha \dot{\alpha}}^\mu \;$being the Pauli matrices.

Moreover, as we shall see in Sect.IV (see eq.(\ref{motion})), the
decomposition (\ref{ss-dec}) will lead to an algebraic structure which will
close on-shell, \textit{i.e. }on the equations of motion of the standard $%
N=1 $ Yang-Mills theory in superspace. In other words, we shall work without
introducing the so called antifields which are known not to contribute to
anomalies and nontrivial counterterms \cite{Barnich}.

We have now to remark that in the last few years the study of the BRS
consistency condition for supersymmetric $N\geq 1$ models has been
undertaken by several authors \cite{D, B, W, M} who faced and solved aspects
which were still open. We mention, for instance, the results concerning the
potential existence of susy anomalies in presence of nonstandard constrained
multiplets \cite{D, B}, and the very useful possibility of performing a
purely algebraic regularization independent analysis of the susy gauge
theories in a fully off-shell version of the Wess-Zumino gauge \cite{W, M}.
In particular, this last result has allowed for a simple discussion of the
renormalization of models with extended supersymmetry (\textit{i.e. }$N\geq
2 $), yielding an algebraic proof of the ultraviolet finiteness of the $N=4$
gauge theories \cite{W}.

However, although much is already known about the BRS cohomology of the $N=1$
gauge theories, we emphasize here that the decomposition formulas (\ref
{ss-dec}) will allow us to cast both the susy algebra and the supersymmetric
BRS transformations into a unique equation which, in complete analogy with
the nonsupersymmetric case, takes the form of a generalized zero curvature
condition. Moreover, by means of this zero curvature equation, we shall be
able to derive the full set of superspace descent equations for the
invariant action and for the $U(1)$ axial anomaly from a unique equation of
the type 
\begin{equation}
\widetilde{d}\;\widetilde{\omega }\;=\;0\;,  \label{nova}
\end{equation}
$\;\widetilde{\omega }$ being a suitable superspace cocycle and $\;%
\widetilde{d}$ the generalized nilpotent operator entering the zero
curvature condition. In addition, a modified version of the eq.(\ref{nova})
will allow to include also the more complex case of the $N=1$ supersymmetric
gauge anomaly, improving thus our understanding of the well known
nonpolynomial character of this anomaly.

The zero curvature equation and the related possibility of collecting the
superspace descent equations into a unique condition represent the main
results of this paper. Their relevance is due to the fact that, besides the
possibility of recovering the solutions of the BRS consistency condition in
a simple way, they provide an interesting pure geometrical framework in $N=1$
superspace.

The work is organized as follows. In Section II we introduce the general
notations and we discuss the supersymmetric decomposition (\ref{ss-dec}).
Section III is devoted to the analysis of the algebraic relations entailed by
the operators $\zeta _\alpha \;$and $\overline{\zeta }_{\dot{\alpha}}$. In
Section IV we present the zero cuvature formulation of the superspace BRS
transformations and of the descent equations corresponding to the invariant
super Yang-Mills lagrangian. In Section V we discuss the descent equations
for the superspace version of the U(1) axial anomaly. Section VI deals with
the case of the supersymmetric chiral gauge anomaly appearing in the quantum
extension of the supersymmetric Slavnov-Taylor identity. In order to make
the paper self contained the final Appendices A, B and C collect a short
summary of the main results concerning the Yang-Mills superspace BRS
cohomology as well as the solution of certain equations relevant for the
superspace version of the descent equations.

\section{General Notations and Decomposition Formulas}

In order to present the general algebraic set up let us begin by fixing the
notations\footnote{%
The superspace conventions used here are those of ref. \cite{pigbook}.}. We
shall work in a four dimensional space-time with N=1 supersymmetry. The
superfield content which will be used throughout is the standard set of the
superfields of the pure N=1 super Yang-Mills theories, \textit{i.e.} the
vector superfield $\phi $ and the gauge superconnections $\varphi _\alpha $
and ${\overline{\varphi }}_{\dot{\alpha}}\,$. They are defined as $\,$ 
\begin{equation}
\begin{array}{l}
\varphi _\alpha \,\equiv \,e^{-\phi }D_\alpha e^\phi \;,\;\;\;\;\;\;\;\;{%
\overline{\varphi }}_{\dot{\alpha}}\,\equiv \,e^\phi \overline{D}_{\dot{%
\alpha}}e^{-\phi }\,,
\end{array}
\label{conec}
\end{equation}
where $D_\alpha \;$and $\overline{D}_{\dot{\alpha}}\;$are the usual
supersymmetric derivatives: 
\begin{equation}
\begin{array}{c}
\left\{ D_\alpha ,D_\beta \right\} \;=\;\left\{ \overline{D}_{\dot{\alpha}},%
\overline{D}_{\dot{\beta}}\right\} \;=\;0\;, \\ 
D_\alpha \,\overline{D}_{\dot{\alpha}}\,+\,\overline{D}_{\dot{\alpha}%
}\,D_\alpha \,=\,2i\sigma _{\alpha \dot{\alpha}}^\mu \;\partial _\mu \;\,.
\end{array}
\label{susy-algebra}
\end{equation}
Introducing now the chiral and antichiral Faddeev-Popov ghosts $c$ and $%
\overline{c}$ 
\[
\overline{D}_{\dot{\alpha}}\;c=D_\alpha \;\overline{c}\;=\;0\,\;, 
\]
for the superspace nilpotent BRS transformations one has 
\begin{equation}
\begin{array}{l}
se^\phi \;=\;e^\phi c\;-\;\overline{c}e^\phi \,, \\ 
sc\;=\;-c^2\,, \\ 
s\overline{c}\;=\;\;-\overline{c}^2\,, \\ 
s\varphi _\alpha \;=\;-D_\alpha c\;-\;\{c,\varphi _\alpha \}\,, \\ 
s\overline{\varphi }_{\dot{\alpha}}\;=\;-\overline{D}_{\dot{\alpha}}%
\overline{c}\;\;-\;\{\overline{c},\overline{\varphi }_{\dot{\alpha}}\}\,.
\end{array}
\label{brs}
\end{equation}
and 
\[
\left\{ s,D_\alpha \right\} \;=\;\left\{ s,\overline{D}_{\dot{\alpha}%
}\right\} \;=\;0\;.\; 
\]
Let us also give, for further use, the BRS transformations of the chiral and
antichiral superfield strengths $F_\alpha $ and $\overline{F}_{\dot{\alpha}}$

\begin{equation}
\begin{array}{l}
F_\alpha \equiv \;{\overline{D}}^2\varphi _\alpha \;,\;\;\;\;\;\;\;\;\;\;\;%
\overline{D}_{\dot{\alpha}}\;F_\alpha =0\;, \\ 
\overline{F}_{\dot{\alpha}}\equiv \;{D}^2{\overline{\varphi }}_{\dot{\alpha}%
}\,\;,\;\;\;\;\;\;\;\;\;\;\;D_\alpha \;\overline{F}_{\dot{\alpha}}=0\;, \\ 
sF_\alpha \;=\;-\{c,F_\alpha \}\,\;,\;\;\;\;\;s\overline{F}_{\dot{\alpha}%
}\;=\;-\{\bar{c},\overline{F}_{\dot{\alpha}}\}\,.
\end{array}
\label{brs-F}
\end{equation}
The quantum numbers, \textit{i.e.} the dimensions, the ghost numbers and the 
$\mathcal{R}$-weights of all the fields are assigned as follows

\begin{table}[hbt]
\centering

%\begin{center}
%$\stackrel{}{\stackunder{\stackunder{}{\mathbf{Table\, 1}}}{\stackrel{%
%\mathbf{R-weights,\, dim. \, and\, ghost\, numb.}\text{\textrm{\ }}}{%
%\TeXButton{mamonas}
%{  
 \begin{tabular}{|c|c|c|c|c|c|c|c|c|c|c|} \hline
        &  $s$ & $D_{\alpha}$ & ${\overline{D}}_{\dot \alpha}$ & $\phi$  & $c$ & $\overline{c}$ & ${\varphi}_{\alpha}$  & ${\overline{\varphi}}_{\dot \alpha}$ & $F_{\alpha}$  & ${\overline{F}}_{\dot \alpha}$  \\ \hline
      $dim$  &  $0$  &  $\frac{1}{2}$ & $\frac{1}{2}$ & $0$ & $0$ & $0$  &  $\frac{1}{2}$ & $\frac{1}{2}$  &  $\frac{3}{2}$  &  $\frac{3}{2}$ \\
    $N_{g}$ &  $1$  &  $0$ & $0$ & $0$ & $1$ & $1$ &  $0$ & $0$ &  $0$ & $0$ \\
     ${\cal R}$      &  $0$  &  $1$ & $-1$ & $0$ & $0$ & $0$  &  $1$ & $-1$  &  $-1$ & $1$  \\ \hline
 \end{tabular}
%}}}}
%\cdot $
%\end{center}

\caption[t1]{Dim.,\, ghost\, numb.\, and\, R-weights.}
\label{mamonas}
\end{table}

The fields will be treated as commuting or anticommuting according to the
fact that their total degree, here chosen to be the sum of the ghost number
and of the spinorial indices, is even or odd. Otherwise stated all the
fields are Lie-algebra valued, the gauge group $\mathcal{G}$ \ being assumed
to be a semisimple Lie group with antihermitian generators $T^a$.$\;$

The set of fields $\left( c,\overline{c},\phi ,\varphi _\alpha ,\overline{%
\varphi }_{\dot{\alpha}}\right) $ and their covariant derivatives will
define therefore the basic local space for studying the superspace descent
equations. Let us also observe that due to the fact that $D,\overline{D}$
have dimension $\frac 12$, the number of covariant derivatives\ turns out to
be limited by power counting requirements. For instance, as we shall see in
the explicit examples considered in the next sections, the analysis of the
superspace consistency condition\ for both the U(1) axial anomaly and the
gauge anomaly requires the use of local formal power series\ in the
variables $\left( c,\overline{c},\phi ,\varphi _\alpha ,\overline{\varphi }_{%
\dot{\alpha}}\right) \;$of dimension 2. We recall here that the non
polynomial character of certain N=1 superspace expressions is due to the
fact that the vector superfield $\phi $ is dimensionless. Finally, whenever
the space time derivatives $\partial _\mu $ appear they are meant to be
replaced by the covariant derivatives $D,\overline{D}$, according to the
supersymmetric algebra $\;$(\ref{susy-algebra}).

Let us introduce now the two operators $\zeta _\alpha \;$and $\overline{%
\zeta }_{\dot{\alpha}}$ of ghost number -1, defined by 
\begin{equation}
\begin{array}{c}
\begin{array}{lcl}
\zeta _\alpha c\;=\;\varphi _\alpha \,, & \;\; & \overline{\zeta }_{\dot{%
\alpha}}\overline{c}\;=\;\overline{\varphi }_{\dot{\alpha}}\,,
\end{array}
\\ 
\zeta _\alpha \overline{c}\;=\;\overline{\zeta }_{\dot{\alpha}}c\;=\;\zeta
_\alpha \phi \;=\;\overline{\zeta }_{\dot{\alpha}}\phi \;=0\;,\,\; \\ 
\;\zeta _\alpha \;\varphi _\beta \;\;=\;\overline{\zeta }_{\dot{\alpha}%
}\;\varphi _\beta \;=\;0\;.
\end{array}
\label{delta definition}
\end{equation}
It is almost immediate thus to check that they are of total degree zero and
that they obey the following algebraic relations 
\begin{equation}
\begin{array}{l}
\left[ \zeta _\alpha ,s\right] \;=\;D_\alpha \,, \\ 
\left[ \overline{\zeta }_{\dot{\alpha}},s\right] \;=\;\overline{D}_{\dot{%
\alpha}}\,,\; \\ 
\left[ \zeta _\alpha ,\zeta _\beta \right] \;=\;\left[ \zeta _\alpha ,%
\overline{\zeta }_{\dot{\beta}}\right] \;=\;\left[ \overline{\zeta }_{\dot{%
\alpha}},\overline{\zeta }_{\dot{\beta}}\right] \;=\;0\;,
\end{array}
\label{super-dec}
\end{equation}
yielding\textit{\ }then the supersymmetric decomposition (\ref{ss-dec}) we
are looking for. As we shall see later on the operators $\zeta _\alpha \;$%
and $\overline{\zeta }_{\dot{\alpha}}$ will turn out to be very useful in
order to solve the superspace descent equations.
Let us focus, for the time being, on the anlysis of the consequences stemming from 
the equations (\ref{super-dec}). 

\section{Algebraic Relations}

To study the algebra entailed by the two operators $\zeta _\alpha \;$and $%
\overline{\zeta }_{\dot{\alpha}}$ let us first observe that they do not
commute with the supersymmetric covariant derivatives $D$,$\;\overline{D}$.
Instead as one can easily check by using the equations (\ref
{delta
definition}) we have, in complete analogy with the nonsupersymmetric
case \cite{Sorella}, 
\begin{eqnarray}
\left[ \overline{\zeta }_{\dot{\beta}},D_\alpha \right] \;\; &=&\;\left[
\zeta _\alpha ,\overline{D}_{\dot{\beta}}\right] \;=\;-\;G_{\alpha \dot{\beta%
}}\,,  \label{Gdef} \\
\left[ \overline{\zeta }_{\dot{\alpha}},\overline{D}_{\dot{\beta}}\right]
\;\; &=&\;\left[ \zeta _\alpha ,D_\beta \right] \;=\;0
\end{eqnarray}
where the new operator $\;G_{\alpha \dot{\beta}}$ has negative ghost number
-1 and acts on the fields as 
\begin{equation}
\begin{array}{c}
\begin{array}{lcl}
G_{\alpha \dot{\alpha}}c\;=\;\overline{D}_{\dot{\alpha}}\varphi _\alpha \,,
& \;\; & G_{\alpha \dot{\alpha}}\overline{c}\;=\;D_\alpha \overline{\varphi }%
_{\dot{\alpha}}\,,
\end{array}
\\ 
G_{\alpha \dot{\alpha}}\;\phi \;=G_{\alpha \dot{\alpha}}\;\varphi _\beta
\,\;=G_{\alpha \dot{\alpha}}\;\overline{\varphi }_{\dot{\beta}}=\;0\;,
\end{array}
\label{G-transform}
\end{equation}
and 
\begin{equation}
\begin{array}{c}
\left\{ G_{\alpha \dot{\alpha}},s\right\} \;=\;\left\{ D_\alpha ,\overline{D}%
_{\dot{\alpha}}\right\} \,,\; \\ 
\left[ \zeta _\alpha ,G_{\beta \dot{\beta}}\right] \;=\;\left[ \overline{%
\zeta }_{\dot{\alpha}},G_{\beta \dot{\beta}}\right] \;=\;\left\{ G_{\alpha 
\dot{\alpha}},G_{\beta \dot{\beta}}\right\} \;=\;\;0\;.
\end{array}
\label{G-dec}
\end{equation}
Again, the operator $G_{\alpha \dot{\beta}}$ does not anticommute with the
covariant derivatives $D$,$\;\overline{D}$. It yields in fact 
\begin{equation}
\begin{array}{cc}
\left\{ G_{\alpha \dot{\alpha}},D_\beta \right\} \;=\;-\frac 12\epsilon
_{\alpha \beta }\overline{R}_{\dot{\alpha}}\,, & \left\{ G_{\alpha \dot{%
\alpha}},\overline{D}_{\dot{\beta}}\right\} \;=\;\frac 12\epsilon _{\dot{%
\alpha}\dot{\beta}}R_\alpha \,,
\end{array}
\label{Rdef}
\end{equation}
with $R_\alpha $ and $\overline{R}_{\dot{\alpha}}\;$of ghost number -1 and
defined as 
\begin{equation}
\begin{array}{c}
\begin{array}{ccc}
R_\alpha c\;=\;F_\alpha \,, & \;\; & \overline{R}_{\dot{\alpha}}\overline{c}%
\;=\;\overline{F}_{\dot{\alpha}}\,,
\end{array}
\\ 
\\ 
R_\alpha \overline{c}\;=\;2\overline{D}_{\dot{\alpha}}D_\alpha \overline{%
\varphi }^{\dot{\alpha}}\,+D_\alpha \overline{D}_{\dot{\alpha}}\overline{%
\varphi }^{\dot{\alpha}}\,+\,\left( D_\alpha \overline{\varphi }_{\dot{\alpha%
}}\right) \overline{\varphi }^{\dot{\alpha}}\,+\overline{\varphi }_{\dot{%
\alpha}}\,\left( D_\alpha \overline{\varphi }^{\dot{\alpha}}\right) \,, \\ 
\\ 
\overline{R}_{\dot{\alpha}}c\;=\;2D^\alpha \overline{D}_{\dot{\alpha}%
}\varphi _\alpha \,+\overline{D}_{\dot{\alpha}}D^\alpha \varphi _\alpha
\,+\,\left( \overline{D}_{\dot{\alpha}}\varphi ^\alpha \right) \varphi
_\alpha \,+\varphi ^\alpha \,\left( \overline{D}_{\dot{\alpha}}\varphi
_\alpha \right) \,, \\ 
\\ 
R_\alpha \;\phi =\;R_\alpha \;\varphi _\beta \,=\;R_\alpha \;\overline{%
\varphi }_{\dot{\beta}}=\;R_\alpha \;F_\beta =\;R_\alpha \;\overline{F}_{%
\dot{\beta}}=\;0\,, \\ 
\\ 
\overline{R}_{\dot{\alpha}}\;\phi =\;\overline{R}_{\dot{\alpha}}\;\varphi
_\beta \,=\;\overline{R}_{\dot{\alpha}}\;\overline{\varphi }_{\dot{\beta}}=\;%
\overline{R}_{\dot{\alpha}}\;F_\beta =\;\overline{R}_{\dot{\alpha}}\;%
\overline{F}_{\dot{\beta}}=\;0.
\end{array}
\label{R-action}
\end{equation}
In addition, we have 
\begin{equation}
\begin{array}{c}
\lbrack R_\alpha ,s]\;=\;[R_\alpha ,D_\beta ]\;=\;[R_\alpha ,\overline{D}_{%
\dot{\beta}}]\;=\;[R_\alpha ,G_{\alpha \dot{\beta}}]\;=\;0\;, \\ 
\lbrack R_\alpha ,\zeta _\beta ]\;=\;[R_\alpha ,\overline{\zeta }_{\dot{\beta%
}}]\;=\;[R_\alpha ,R_\beta ]\;=\;[R_\alpha ,\overline{R}_{\dot{\beta}%
}]\;=\;\;0\;.\;
\end{array}
\label{commutators}
\end{equation}
Let us finally display the quantum numbers of the operators entering the
algebraic relations (\ref{super-dec}),\ (\ref{Gdef}),\ (\ref{Rdef})

\begin{table}[hbt]
\centering

%\begin{center}
%$\stackunder{\stackunder{}{\mathbf{Table\, 2}}}{\stackrel{\mathbf{%
%R-weights,\, dim. \, and\, ghost\, numb.}\text{\textrm{\ }}}{%
%\TeXButton{operators}
%{ 
\begin{tabular}{|l|l|l|l|l|l|} \hline
        &  $\zeta^{\alpha}$  & ${\overline{\zeta}}^{\dot \alpha}$ & $G^{\alpha \dot \alpha}$ & $R^{\alpha}$ & ${\overline R}^{\dot \alpha}$ \\ \hline
      $dim$  &  $\frac{1}{2}$ & $\frac{1}{2}$ & $1$ & $\frac{3}{2}$ & $\frac{3}{2}$   \\
    $N_{g}$ &  $-1$  &  $-1$ & $-1$ & $-1$ & $-1$   \\
     ${\cal R}$      &  $1$  &  $-1$ & $0$ & $-1$ & $1$  \\ \hline
  \end{tabular}
% }}}$
%\end{center}

\caption[t1]{ Dim.,\, ghost\, numb.\, and\, R-weights.}
\label{operators }
\end{table}

In the next Section it will be shown how the operators in the Table 2 can be combined 
into a unique generalized operator by means of the introduction of a set of global 
parameters. These parameters will be required to fulfill a certain number of 
suitable
conditions (see eqs.(\ref{basis}) in the next Sect.), which will project 
the algebra (\ref{super-dec}),\ (\ref{Gdef}),\ (\ref{Rdef}) on the equations 
of motion of N=1 super 
Yang-Mills and will allow us 
to cast the BRS transformations of the superfields in the form of a 
zero curvature condition in superspace. 
In particular, the introduction of the aforementioned global parameters will have the 
effect of realizing  the decomposition (\ref{super-dec}) on all the elementary 
superfields $\left( c,\overline{c},\phi ,\varphi
_\alpha ,\overline{\varphi }_{\dot{\alpha}}\right) $ and their covariant derivatives 
on shell, {\it i.e.}, 
modulo the equations of motion. 
Therefore, we can assume as the basic functional space for the forthcoming  analysis 
that of the polynomials in the elementary superfields and their covariant
derivatives bounded by dimension two. This limitation comes directly by
superspace power-counting considerations. Moreover, all possible nontrivial
counterterms and anomalies allowed by the power-counting will be
included. 

\section{The Zero Curvature Condition}

Having characterized all the relevant operators entailed by the consistency
of the supersymmetric decomposition (\ref{super-dec}), let us pay attention
to the geometrical aspects of the algebraic relations so far obtained. To
this purpose it is useful to introduce a set of global parameters $e^\alpha $%
, $\overline{e}^{\dot{\alpha}}$ and $\widetilde{e}^{\alpha \dot{\alpha}}$,
naturally associated to the operators $\zeta _\alpha $, $\overline{\zeta }_{%
\dot{\alpha}}$ and $G_{\alpha \dot{\beta}}$, of ghost number one and obeying
the relations 
\begin{equation}
\begin{array}{c}
e^\alpha \;e^\beta \;=\;\overline{e}^{\dot{\alpha}}\;\overline{e}^{\dot{\beta%
}}\;=\;\widetilde{e}^{\alpha \dot{\alpha}}\;\widetilde{e}^{\beta \dot{\beta}%
}\;=\;0\;, \\ 
\left[ e^\alpha ,\overline{e}^{\dot{\beta}}\right] \;=\;\left[ e^\alpha ,%
\widetilde{e}^{\beta \dot{\beta}}\right] \;=\;\left[ \widetilde{e}^{\alpha 
\dot{\alpha}},\overline{e}^{\dot{\beta}}\right] \;=\;0\;,
\end{array}
\label{susy-base}
\end{equation}

\begin{table}[hbt]
\centering

%\begin{center}
%$\stackunder{\mathbf{Table \,3 }}{\stackrel{\mathbf{R-weights,\, dim. \,
%and\, ghost\, numb.}}{\TeXButton{ob}
%{
  \begin{tabular}{|l|l|l|l|} \hline
       &   $e^{\alpha}$ & ${\overline e}^{\dot \alpha}$ & ${\tilde e}^{\alpha \dot \alpha}$ \\ \hline
     $dim$  & $ -\frac{1}{2}$ & $-\frac{1}{2}$ & $-1$  \\
   $N_{g}$  & $1$ & $1$ & $1$  \\
    ${\cal R}$  & $0$ & $0$ & $0$ \\ \hline
 \end{tabular}
%}}}$
%\end{center}

\caption[t1]{ Dim.,\, ghost\, numb.\, and\, R-weights.}
\label{ob}
\end{table}

In addition, the global parameters ($e^\alpha $, $\overline{e}^{\dot{\alpha}%
},$ $\widetilde{e}^{\alpha \dot{\alpha}}$) will be required to obey the
following conditions (see also Appendix D)

\begin{equation}
\begin{tabular}{l}
$e^\alpha \;\widetilde{e}^{\beta \dot{\alpha}}\;=-\frac 12\epsilon ^{\alpha
\beta }\;e^\gamma \;\widetilde{e}_\gamma ^{\dot{\alpha}}\;,$ \\ 
$\widetilde{e}^{\alpha \dot{\alpha}}\;\overline{e}^{\dot{\beta}}\;=\;\frac
12\epsilon ^{\dot{\alpha}\dot{\beta}}\;\widetilde{e}_{\dot{\gamma}}^\alpha \;%
\overline{e}^{\dot{\gamma}}\;,$ \\ 
$e^\alpha \;\widetilde{e}^{\beta \dot{\alpha}}\;\overline{e}^{\dot{\beta}%
}\;=-\frac 14\epsilon ^{\alpha \beta }\epsilon ^{\dot{\alpha}\dot{\beta}%
}\;e^\gamma \;\widetilde{e}_{\gamma \dot{\gamma}}\;\overline{e}^{\dot{\gamma}%
}\;,$%
\end{tabular}
\label{basis}
\end{equation}
fixing the symmetry properties of the product of two parameters with respect
to their spinorial indices. Defining now the nilpotent dimensionless
operators $\zeta $, $\overline{\zeta }$\ and $G$ as

\[
\zeta \;=\;\zeta ^\alpha \;e_\alpha \;,\;\;\;\;\overline{\zeta }\;=\;%
\overline{\zeta }_{\dot{\alpha}}\;\overline{e}^{\dot{\alpha}%
}\;,\;\;\;\;G\;=\;G_{\dot{\alpha}}^\alpha \;\widetilde{e}_\alpha ^{\dot{%
\alpha}}\;, 
\]
it is straightforward to verify that they have zero ghost number and $%
\mathcal{R}$ weight respectively 1, -1, 0, and that the subalgebra generated
by $\zeta _\alpha $, $\overline{\zeta }_{\dot{\alpha}}$ and $G_{\alpha \dot{%
\beta}}$, \textit{i.e.}

\[
\begin{array}{c}
\left[ \zeta _\alpha ,\zeta _\beta \right] \;=\;\left[ \zeta _\alpha ,%
\overline{\zeta }_{\dot{\beta}}\right] \;=\;\left[ \overline{\zeta }_{\dot{%
\alpha}},\overline{\zeta }_{\dot{\beta}}\right] \;=\;0\;, \\ 
\left[ \zeta _\alpha ,G_{\beta \dot{\beta}}\right] \;=\;\left[ \overline{%
\zeta }_{\dot{\alpha}},G_{\beta \dot{\beta}}\right] \;=\;\left\{ G_{\alpha 
\dot{\alpha}},G_{\beta \dot{\beta}}\right\} \;=\;\;0\;,
\end{array}
\]
can be simply rewritten as

\[
\left[ \zeta ,\overline{\zeta }\right] \;=\;\left[ \zeta ,G\right]
\;=\;\left[ \overline{\zeta },G\right] \;=\;0\;. 
\]
Analogously, introducing the nilpotent operators $\widetilde{G}$, $D$, $%
\overline{D}$, $R$, $\overline{R}$, $\partial $, $\widetilde{\partial }$

\begin{equation}
\begin{array}{ccccccc}
\widetilde{G} & = & G_{\dot{\alpha}}^\alpha \;e_\alpha \;\overline{e}^{\dot{%
\alpha}}\;, &  & G & = & G_{\dot{\alpha}}^\alpha \;\widetilde{e}_\alpha ^{%
\dot{\alpha}}\;, \\ 
&  &  &  &  &  &  \\ 
D & = & D^\alpha \;e_\alpha \;, &  & \overline{D} & = & \overline{D}_{\dot{%
\alpha}}\;\overline{e}^{\dot{\alpha}}\;, \\ 
&  &  &  &  &  &  \\ 
R & = & R^\alpha \;\widetilde{e}_{\alpha \dot{\alpha}}\;\overline{e}^{\dot{%
\alpha}}\;, &  & \overline{R} & = & \overline{R}_{\dot{\alpha}}\;e^\alpha \;%
\widetilde{e}_\alpha ^{\dot{\alpha}}\;, \\ 
&  &  &  &  &  &  \\ 
\widetilde{\partial } & = & \left\{ D^\alpha ,\overline{D}_{\dot{\alpha}%
}\right\} \;e_\alpha \;\overline{e}^{\dot{\alpha}}\;, &  & \partial & = & 
\left\{ D^\alpha ,\overline{D}_{\dot{\alpha}}\right\} \;\widetilde{e}_\alpha
^{\dot{\alpha}}\;,\;
\end{array}
\label{conv}
\end{equation}
it is immediate to check that all the algebraic relations and field
transformations of eqs.(\ref{delta definition})-eqs.(\ref{commutators}) may
be cast into the following free index notation

\begin{equation}
\begin{array}{cccccccccccc}
\left[ \zeta ,s\right] & = & D\;, & \left[ \overline{\zeta },s\right] & = & 
\overline{D}\;, & \left\{ \widetilde{G},s\right\} & = & \widetilde{\partial }%
\;, & \left[ G,s\right] & = & \partial \;, \\ 
&  &  &  &  &  &  &  &  &  &  &  \\ 
\left\{ s,D\right\} & = & 0\;, & \left\{ s,\overline{D}\right\} & = & 0\;, & 
\left[ s,\widetilde{\partial }\right] & = & 0\;, & \left\{ s,\partial
\right\} & = & 0\;, \\ 
&  &  &  &  &  &  &  &  &  &  &  \\ 
\left[ D,\widetilde{\partial }\right] & = & 0\;, & \left[ \overline{D},%
\widetilde{\partial }\right] & = & 0\;, & \left\{ D,\partial \right\} & = & 
0\;, & \left\{ \overline{D},\partial \right\} & = & 0\;, \\ 
&  &  &  &  &  &  &  &  &  &  &  \\ 
\left[ D,\zeta \right] & = & 0\;, & \left[ \overline{D},\overline{\zeta }%
\right] & = & 0\;, & \left[ \overline{D},\zeta \right] & = & \widetilde{G}\;,
& \left[ D,\overline{\zeta }\right] & = & \widetilde{G}\;, \\ 
&  &  &  &  &  &  &  &  &  &  &  \\ 
\left[ \partial ,\widetilde{\partial }\right] & = & 0\;, & \left[ G,%
\widetilde{G}\right] & = & 0\;, & \left[ \zeta ,\widetilde{G}\right] & = & 
0\;, & \left[ \overline{\zeta },\widetilde{G}\right] & = & 0\;, \\ 
&  &  &  &  &  &  &  &  &  &  &  \\ 
\left[ G,\partial \right] & = & 0\;, & \left[ \widetilde{G},\widetilde{%
\partial }\right] & = & 0\;, & \left[ G,\widetilde{\partial }\right] & = & 
0\;, & \left\{ \widetilde{G},\partial \right\} & = & 0\;,
\end{array}
\label{free-algebra}
\end{equation}
\[
\begin{array}{cccccccccccc}
\left\{ \widetilde{G},D\right\} & = & 0\;, & \left\{ \widetilde{G},\overline{%
D}\right\} & = & 0\;, & 2\left[ D,G\right] & = & \overline{R}\;, & 2\left[ G,%
\overline{D}\right] & = & R\;, \\ 
&  &  &  &  &  &  &  &  &  &  &  \\ 
\left[ \zeta ,\widetilde{\partial }\right] & = & 0\;, & \left[ \overline{%
\zeta },\widetilde{\partial }\right] & = & 0\;, & 2\left[ \zeta ,\partial
\right] & = & \overline{R}\;, & 2\left[ \partial ,\overline{\zeta }\right] & 
= & R\;, \\ 
&  &  &  &  &  &  &  &  &  &  &  \\ 
\left[ \zeta ,R\right] & = & 0\;, & \left[ \zeta ,\overline{R}\right] & = & 
0\;, & \left[ \overline{\zeta },R\right] & = & 0\;, & \left[ \overline{\zeta 
},\overline{R}\right] & = & 0\;, \\ 
&  &  &  &  &  &  &  &  &  &  &  \\ 
\left[ R,\widetilde{\partial }\right] & = & 0\;, & \left[ \overline{R},%
\widetilde{\partial }\right] & = & 0\;, & \left\{ R,\partial \right\} & = & 
0\;, & \left\{ \overline{R},\partial \right\} & = & 0\;, \\ 
&  &  &  &  &  &  &  &  &  &  &  \\ 
\left\{ D,R\right\} & = & 0\;, & \left\{ \overline{D},R\right\} & = & 0\;, & 
\left\{ D,\overline{R}\right\} & = & 0\;, & \left\{ \overline{D},\overline{R}%
\right\} & = & 0\;, \\ 
&  &  &  &  &  &  &  &  &  &  &  \\ 
\left\{ \widetilde{G},R\right\} & = & 0\;, & \left\{ \widetilde{G},\overline{%
R}\right\} & = & 0\;, & \left[ G,R\right] & = & 0\;, & \left[ G,\overline{R}%
\right] & = & 0\;, \\ 
&  &  &  &  &  &  &  &  &  &  &  \\ 
\left\{ s,R\right\} & = & 0\;, & \left\{ s,\overline{R}\right\} & = & 0\;, & 
\left\{ R,\overline{R}\right\} & = & 0\;, &  &  & 
\end{array}
\]
Let us proceed now by showing that, as announced in the introduction, the
supersymmetric BRS transformations (\ref{brs}), (\ref{brs-F}) can be
obtained by means of a generalized zero curvature condition. To this aim let
us introduce the operator $\delta $

\begin{equation}
\;\delta \;=\;\zeta \;+\;\overline{\zeta }\;\;-\;G\;,  \label{susy-dec}
\end{equation}
from which one easily obtains the following decomposition 
\[
\left[ s,\delta \right] \;=\;-D\;-\;\overline{D}\;-\;\partial \;. 
\]
Defining now the $\delta $-transform of the BRS operator $s$ as 
\begin{equation}
\widetilde{d}\;=\;e^\delta \;s\;e^{-\delta }\;,  \label{dtilde}
\end{equation}
one gets 
\begin{eqnarray}
\widetilde{d}\; &=&\;s\;+\;D\;+\;\overline{D}\;+\;\;\partial \;-\;\widetilde{%
G}\;+\;\frac 12\;\overline{R}\;-\frac 12R\;,  \label{dtilde1} \\
\widetilde{d}\;\widetilde{d}\; &=&\;0\;,  \nonumber
\end{eqnarray}
so that, calling $\widetilde{A}\;$and $\widetilde{\overline{A}}$ the $\delta 
$-transform of the chiral and antichiral ghosts $(c,\overline{c})$ $\;$ 
\begin{equation}
\widetilde{A}\;=\;e^\delta \;c\;=\;c\;+\;\varphi \;+\;\overline{D}\;\varphi
\;,\;\;\;\;\;\;\;\;\;\varphi \;=\;\varphi ^\alpha e_\alpha \;,
\label{Atilde-con}
\end{equation}
\begin{equation}
\widetilde{\overline{A}}\;=\;e^\delta \;\overline{c}\;=\;\overline{c}\;+\;%
\overline{\varphi }\;+\;D\;\overline{\varphi }\;,\;\;\;\;\;\;\;\;\;\overline{%
\varphi }\;=\;\overline{\varphi }_{\dot{\alpha}}\overline{e}^{\dot{\alpha}%
}\;,  \label{Abartilde-con}
\end{equation}
it follows that the BRS transformations of $\left( c,\overline{c}\right) $
imply the zero curvature equations 
\begin{equation}
e^\delta \;s\;e^{-\delta }\;e^\delta \;c\;=\;-e^\delta
\;c^2\;\Longrightarrow \;\widetilde{d}\;\widetilde{A}\;+\;\widetilde{A}%
^2\;=\;0\;,  \label{zero-curv}
\end{equation}
and 
\begin{equation}
e^\delta \;s\;e^{-\delta }\;e^\delta \;\overline{c}\;=\;-e^\delta \;%
\overline{c}^2\;\Longrightarrow \;\widetilde{d}\;\widetilde{\overline{A}}%
\;+\;\widetilde{\overline{A}}^2\;=\;0\;.  \label{zero-curv-bar}
\end{equation}
Equations (\ref{zero-curv}) and (\ref{zero-curv-bar}) are easily checked to
reproduce all the BRS transformations (\ref{brs}), (\ref{brs-F}) as well as
the whole set of the equations (\ref{G-transform})-(\ref{R-action}). One
sees thus that, in complete analogy with the nonsupersymmetric case \cite
{zero-curv}, the zero curvature equations (\ref{zero-curv}) and (\ref
{zero-curv-bar}) deeply rely on the existence of the operators $\zeta
_\alpha $ and $\overline{\zeta }_{\dot{\alpha}}$. Let us underline here that
the nilpotent operator $\widetilde{d}$ in eq.(\ref{dtilde1}) will play a
rather important role in the discussion of the superspace descent equations.
For instance, as we shall see explicitly in the example given in the next
subsection, it turns out that the superspace descent equations corresponding
to the BRS invariant counterterms can be remarkably obtained from the single
equation 
\begin{equation}
\widetilde{d}\;\widetilde{\omega }\;=\;0\;,  \label{desc1}
\end{equation}
where $\widetilde{\omega }$ is an appropriate cocycle of dimension zero and
ghost number three, whose components are the superspace field polynomials of
the Taylor expansion of $\widetilde{\omega }$ in the global parameters ($%
e^\alpha $, $\overline{e}^{\dot{\alpha}}$, $\widetilde{e}^{\alpha \dot{\alpha%
}}$). Equation (\ref{desc1}) can also be applied to characterize the descent
equations of the U(1) anomaly. In Sect.VI we shall see that a slight
modification of the eq.(\ref{desc1}) will allow to treat the case of the
Yang-Mills gauge anomaly as well. In all these cases the components of $%
\widetilde{\omega }$ will not exceed dimension two, this dimension being
taken as the upper limit of our superspace analysis of the descent
equations. In other words in what follows we shall limit ourselves to the
study of the solutions of the superspace descent equations in the space of
local functionals with dimension less or equal to two. In particular,
according to the Table 1, this implies that the maximum number of covariant
derivatives $D,\overline{D}$ present in each component of $\widetilde{\omega 
}$ is four.

Let us conclude this section with the following important remark. Being
interested in the descent equations involving superspace functionals of
dimension less or equal to two, we should have checked the closure of the
algebra (\ref{free-algebra}) built up by the operators $(s,\zeta ,\overline{%
\zeta },G,\widetilde{G},D,\overline{D},R,\overline{R},\partial ,\widetilde{%
\partial })$ on all the fields and their covariant derivatives up to
reaching dimension two. It is not difficult to convince oneself that
actually there is a breakdown of the closure of this algebra in the highest
level of dimension two. However, as it usually happens in supersymmetry, the
breaking terms turn out to be nothing but the equations of motion
corresponding to the pure N=1 susy Yang-Mills action, implying thus an
on-shell closure of the algebra. Evaluating in fact the commutator between
the operators $\zeta $ and $s$ on the superfield strength $F$ one gets 
\begin{equation}
\left[ \zeta \;,\;s\right] \;F\;=\;-\;\left[ \varphi \;,\;F\right] \;.
\label{on-shell}
\end{equation}
The right hand side of the equation (\ref{on-shell}) can be rewritten as 
\[
\left[ \zeta \;,\;s\right] \;F\;=D\;F\;-\;(\;D\;F\;+\;\left[ \varphi
\;,\;F\right] \;)\;, 
\]
so that, recalling that 
\begin{equation}
D\;F\;\;+\;\left[ \varphi \;,\;F\right] \;=-\frac 12e^\gamma \widetilde{e}%
_{\gamma \dot{\gamma}}\overline{e}^{\dot{\gamma}}\;\left( D^\alpha
\;F_\alpha \;+\;\left\{ \varphi ^\alpha \;,\;F_\alpha \right\} \right)
\;=\;0\;,  \label{YM-eq-motion}
\end{equation}
are precisely the equations of motion of the pure N=1 susy Yang-Mills
action, one obtains

\begin{equation}
\left[ \zeta \;,\;s\right] \;F\;=D\;F\;-\;\mathrm{eq.\;of\;motion}\;.
\label{motion}
\end{equation}

It is worth underlining here that the on-shell closure of the algebra relies 
precisely on the introduction of the global parameters $%
e^\alpha $, $\overline{e}^{\dot{\alpha}}$, $\widetilde{e}^{\alpha \dot{\alpha%
}}$ and on the relations (\ref{basis}). Nevertheless, this on-shell closure 
does
not represent a real obstruction in order to solve the superspace
consistency conditions. In fact,  from the eq.(\ref
{YM-eq-motion}) and from the Table 1,  one can observe that the  equations 
of motion of N=1 super 
Yang-Mills 
are of dimension two. 
Therefore, they could eventually contribute only to the
highest level of the descent equations. Rather, the above on-shell
closure  (\ref{free-algebra})  is related to the absence of the so called BRS external
fields ({\it i.e.} the Batalin-Vilkoviski antifields) which are known 
to properly take care of the equations of motion. However, as shown by 
\cite{pig1,pig2,pigbook}, these external fields do not contribute to the
superspace BRS cohomology in the cases considered here of the U(1) chiral
anomaly, of the gauge anomaly as well as of the invariant counterterms. This is
the reason why we have discarded them. In the App.C it will be  shown how the
introduction of an appropriate external field takes care in a simple way of
the Yang-Mills equations of motion, closing thus the algebra off-shell.

\subsection{Nonchiral Descent Equations for the Invariant Action}

In order to apply the supersymmetric decomposition (\ref{super-dec}) to the
analysis of the superspace descent equations, let us begin by considering
the BRS consistency condition corresponding to the nonchiral Yang-Mills
invariant action, \textit{i.e.} 
\begin{equation}
s\int d^4x\;d^2\theta \;d^2\overline{\theta }\;{\cal{L}}
^0\;=\;0\;\Longrightarrow \;s\;
{\cal{L}}
^0\;=\;D^\alpha \;{\cal{L}}%
_\alpha ^1\;+\;\;\overline{D}_{\dot{\alpha}}\;{\cal{L}}^{1\dot{\alpha}}\;,
\label{nonchiral-actionWZ}
\end{equation}
where ${\cal{L}}^0$ is a local power series of dimension two and ghost
number zero. According to what mentioned in the previous section, the full
set of the superspace descent equations characterizing ${\cal{L}}^0$ can
be obtained directly from the generalized equation 
\begin{equation}
\widetilde{d}\;\widetilde{\omega }\;=\;0\;,  \label{desc2}
\end{equation}
with $\widetilde{\omega }\;$ a generalized cocycle of ghost number three and
dimension zero, whose Taylor expansion in the global parameters $(e^\alpha ,%
\widetilde{e}^{\alpha \dot{\alpha}},\overline{e}^{\dot{\alpha}})$ reads 
\begin{equation}
\begin{array}{ccc}
\widetilde{\omega } & = & \;\omega ^3\;+\;\;\omega ^{2\;\alpha }\;e_\alpha
\;+\;\;\overline{\omega }_{\;\dot{\alpha}}^2\;\overline{e}^{\dot{\alpha}%
}\;+\;\widetilde{\omega }_{\;\;\dot{\alpha}}^{2\;\alpha }\;\widetilde{e}%
_\alpha ^{\dot{\alpha}}\;+\;\widetilde{\omega }_{\;\;\dot{\alpha}%
}^{1\;\alpha }\;e_\alpha \;\overline{e}^{\dot{\alpha}} \\ 
&  &  \\ 
&  & +\;\omega ^{1\;\alpha }\;\widetilde{e}_{\alpha \dot{\alpha}}\;\overline{%
e}^{\dot{\alpha}}\;+\;\overline{\omega }_{\;\dot{\alpha}}^1\;e^\alpha \;%
\widetilde{e}_\alpha ^{\dot{\alpha}}\;+\;\omega ^0\;e^\alpha \;\widetilde{e}%
_{\alpha \dot{\alpha}}\;\overline{e}^{\dot{\alpha}}\;.
\end{array}
\label{wtilde}
\end{equation}
The coefficients $(\omega ^3,\omega ^{2\;\alpha },\overline{\omega }_{\;\dot{%
\alpha}}^2,\widetilde{\omega }_{\;\;\dot{\alpha}}^{2\;\alpha },\widetilde{%
\omega }_{\;\;\dot{\alpha}}^{1\;\alpha },\omega ^{1\;\alpha },\overline{%
\omega }_{\;\dot{\alpha}}^1,\omega ^0)$ are local power series in the
superfields with the following quantum numbers

\begin{table}[hbt]
\centering

%\begin{center}
%$\stackrel{\mathbf{R-weights,\, dim. \, and\, ghost\, numb.}}{%
%\TeXButton{omegas}
%{
 \begin{tabular}{|l|l|l|l|l|l|l|l|l|} \hline
        &  $\omega ^3$  & $\omega ^{2\;\alpha }$ & $\overline{\omega }_{\;\dot \alpha }^2$ & $\widetilde{\omega }_{\;\;\dot \alpha }^{2\;\alpha }$ & $\widetilde{\omega }_{\;\;\dot \alpha }^{1\;\alpha }$ & $\omega ^{1\;\alpha }$ & $\overline{\omega }_{\;\dot \alpha }^1$ & $\omega ^0$\\ \hline
      $dim$  &  $0$ & $\frac{1}{2}$ & $\frac{1}{2}$ & $1$ & $1$ &$\frac{3}{2}$ & $\frac{3}{2}$ & $2$  \\
    $N_{g}$ &  $3$  &  $2$ & $2$ & $2$ & $1$ & $1$ & $1$ & $0$ \\ \hline
  \end{tabular}
% }}$

%$\stackrel{\mathbf{Table \, 4}}{}$
%\end{center}

\caption[t1]{Dim.\, and\, ghost\, numb.}
\label{omegas}
\end{table}

In particular one observes that the coefficient $\;\omega ^0$ in the
expression (\ref{wtilde}) has the same dimension of the invariant action we
are looking for, justifying thus the choice of the quantum numbers of $%
\widetilde{\omega }$ in eq.(\ref{desc2}).

The generalized condition (\ref{desc2}) is easily worked out and yields the
following set of equations 
\begin{equation}
\begin{array}{lll}
s\;\omega ^0 & = & -\frac 12D^\alpha \omega _{\;\;\alpha }^{1\;}\;+\frac 12%
\overline{D}_{\dot{\alpha}}\overline{\omega }_{\;}^{1\;\dot{\alpha}}\;+\frac
14\overline{R}_{\dot{\alpha}}\overline{\omega }^{2\;\dot{\alpha}}\;+\frac
14R^\alpha \omega _{\;\alpha }^2\; \\ 
&  &  \\ 
&  & \;\;\;\;\;-\frac 14\left\{ D^\alpha ,\overline{D}_{\dot{\alpha}%
}\right\} \widetilde{\omega }_{\;\;\alpha }^{1\;\dot{\alpha}}\;-\frac 14G_{%
\dot{\alpha}}^\alpha \;\widetilde{\omega }_{\;\;\alpha }^{2\;\dot{\alpha}}\;,
\\ 
&  &  \\ 
s\;\overline{\omega }_{\;\dot{\alpha}}^1\; & = & -\frac 12\;\left\{ D^\alpha
,\overline{D}_{\dot{\alpha}}\right\} \;\omega _{\;\alpha }^{2\;}\;\;-\frac
12\;D^\alpha \;\widetilde{\omega }_{\;\alpha \dot{\alpha}}^2\;+\frac 12%
\overline{R}_{\dot{\alpha}}\;\omega ^3\;, \\ 
&  &  \\ 
s\;\widetilde{\omega }_{\;\;\dot{\alpha}}^{1\;\alpha }\; & = & -D^\alpha \;%
\overline{\omega }_{\;\dot{\alpha}}^2\;\;-\overline{D}_{\dot{\alpha}%
}\;\omega ^{2\;\alpha }\;\;+\;G_{\dot{\alpha}}^\alpha \;\omega ^3\;, \\ 
&  &  \\ 
s\;\omega ^{1\;\alpha }\; & = & \frac 12\;\left\{ D^\alpha ,\overline{D}_{%
\dot{\alpha}}\right\} \;\overline{\omega }^{2\;\dot{\alpha}}\;\;+\;\frac 12\;%
\overline{D}_{\dot{\alpha}}\;\widetilde{\omega }^{2\;\alpha \dot{\alpha}%
}\;\;-\;\frac 12R^\alpha \;\omega ^3\;, \\ 
&  &  \\ 
s\;\overline{\omega }_{\;\dot{\alpha}}^2\;\; & = & -\overline{D}_{\dot{\alpha%
}}\;\omega ^3\;, \\ 
&  &  \\ 
s\;\widetilde{\omega }_{\;\;\dot{\alpha}}^{2\;\alpha } & = & \left\{
D^\alpha ,\overline{D}_{\dot{\alpha}}\right\} \;\omega ^3\;, \\ 
&  &  \\ 
s\;\omega ^{2\;\alpha } & = & -D^\alpha \;\omega ^3\;, \\ 
&  &  \\ 
s\;\omega ^3\; & = & \;0\;.
\end{array}
\label{desc-count}
\end{equation}
These equations do not yet represent the final version of the superspace
descent equations, due to the presence of the operators $(G_{\alpha \dot{%
\alpha}},R_\alpha ,\overline{R}_{\dot{\alpha}})$ in their right hand sides.
However we shall prove that these undesired terms can be rewritten as pure
BRS cocycles or as total superspace derivatives, meaning that they can be
eliminated by means of a redefinition of the $\omega $ 's cocycles entering
the equations (\ref{desc-count}). Let us first observe that a particular
solution of the tower (\ref{desc-count}) can be fully expressed in terms of
the BRS invariant cocycle $\omega ^3$. In fact owing to the zero curvature
equations (\ref{dtilde}), (\ref{zero-curv}) and (\ref{zero-curv-bar}) it is
apparent that the system (\ref{desc-count}) is solved by 
\begin{equation}
\widetilde{\omega }\;=\;e^\delta \;\omega ^3\;,  \label{wtilde-sol}
\end{equation}
which when written in components yields the following expressions 
\begin{equation}
\begin{array}{lll}
\omega ^{2\;\alpha } & = & \;\;\zeta ^\alpha \;\omega ^3\;, \\ 
&  &  \\ 
\widetilde{\omega }_{\;\;\dot{\alpha}}^{2\;\alpha }\; & = & \;G_{\dot{\alpha}%
}^\alpha \;\omega ^3\;, \\ 
&  &  \\ 
\overline{\omega }_{\;\dot{\alpha}}^2 & = & \;\;\overline{\zeta }_{\dot{%
\alpha}}\;\omega ^3\;, \\ 
&  &  \\ 
\omega ^{1\;\alpha }\; & = & \;\frac 12\;G_{\dot{\alpha}}^\alpha \;\overline{%
\zeta }^{\dot{\alpha}}\;\omega ^3\;, \\ 
&  &  \\ 
\widetilde{\omega }_{\;\;\dot{\alpha}}^{1\;\alpha } & = & \;\zeta ^\alpha \;%
\overline{\zeta }_{\dot{\alpha}}\;\omega ^3\;, \\ 
&  &  \\ 
\overline{\omega }_{\;\dot{\alpha}}^1\; & = & \;-\frac 12\;G_{\dot{\alpha}%
}^\alpha \;\zeta _\alpha \;\omega ^3\;, \\ 
&  &  \\ 
\omega ^0 & = & \;\;\frac 14\;\zeta ^\alpha \;G_{\alpha \dot{\alpha}}\;%
\overline{\zeta }^{\dot{\alpha}}\;\omega ^3\;.
\end{array}
\label{sol-count}
\end{equation}
In particular, from the results on the superspace BRS cohomology \cite
{pig2,pig3,porrati}(see the App.B), it turns out that the most general form
for $\omega ^3$ can be identified with the invariant ghost monomial 
\begin{equation}
\;Tr\left( \frac{c^3}3\right) \;,  \label{c3}
\end{equation}
which, of course, is determined modulo a trivial exact BRS cocycle.
Recalling then (App.B) that the difference $\left( Tr\;c^3-Tr\;\overline{c}%
^3\;\right) $ is cohomologically trivial, \textit{i.e.} 
\[
Tr\;c^3\;-\;Tr\;\overline{c}^3\;=\;s\;(...)\;, 
\]
we can choose for $\omega ^3$ the following symmetric expression\footnote{%
One should observe that due to the anti-hermiticity property of the group
generators $T^a$ the cocycle $(Tr\;c^3+\;Tr\;\overline{c}^3)$ is real.} 
\begin{equation}
\omega ^3\;=\;Tr\left( \frac{c^3}3\right) \;+\;Tr\left( \frac{\overline{c}^3}%
3\right) \;.  \label{ghost-cocycle-c3}
\end{equation}
On the other hand it is easily established that all the terms $R^\alpha
\;\omega ^3$, $\overline{R}_{\dot{\alpha}}\;\omega ^3$, $R^\alpha \;\omega
_{\;\alpha }^2$, $\overline{R}_{\dot{\alpha}}\;\overline{\omega }^{2\;\dot{%
\alpha}}$ in the right hand side of eqs.(\ref{desc-count}) are trivial BRS
cocycles. Considering for instance the first term, we have from the eqs.(\ref
{commutators}) 
\begin{equation}
s\;R^\alpha \;\omega ^3\;=\;\;R^\alpha \;s\;\omega ^3\;=\;0\;,\;
\label{R-trivial}
\end{equation}
which implies that $R^\alpha \;\omega ^3$ belongs to the cohomology of $s$
in the sector of ghost number two and dimension one half. Therefore, being
the BRS cohomology empty in this sector, it follows that 
\begin{equation}
R^\alpha \;\omega ^3\;=\;s\;\Lambda ^{1\;\alpha }\;,  \label{Rao3}
\end{equation}
as well as 
\begin{equation}
\overline{R}^{\dot{\alpha}}\;\omega ^3\;=\;s\;\overline{\Lambda }^{1\;\dot{%
\alpha}}\;.  \label{Rbao3}
\end{equation}
In fact, from 
\begin{eqnarray*}
R^\alpha \;Tr\left( \frac{c^3}3\right) \; &=&\;s\;Tr\left( cR^\alpha
c\right) \;=\;s\;Tr\left( cF^\alpha \right) \;, \\
\overline{R}^{\dot{\alpha}}\;Tr\left( \frac{c^3}3\right) \; &=&\;s\;Tr\left(
c\overline{R}^{\dot{\alpha}}c\right) \;,
\end{eqnarray*}
and 
\begin{eqnarray*}
R^\alpha \;Tr\left( \frac{\overline{c}^3}3\right) \; &=&\;s\;Tr\left( 
\overline{c}R^\alpha \overline{c}\right) \;\;, \\
\overline{R}^{\dot{\alpha}}\;Tr\left( \frac{\overline{c}^3}3\right) \;
&=&\;s\;Tr\left( \overline{c}\overline{R}^{\dot{\alpha}}\overline{c}\right)
=\;s\;Tr\left( \overline{c}\overline{F}^{\dot{\alpha}}\right) \;,
\end{eqnarray*}
we have that $\Lambda ^{1\;\alpha }$ and $\overline{\Lambda }^{1\;\dot{\alpha%
}}$ can be identified, modulo trivial terms, with 
\begin{equation}
\Lambda ^{1\;\alpha }\;=\;Tr\left( cF^\alpha \right) \;+Tr\left( \overline{c}%
R^\alpha \overline{c}\right) \;\;,\;\;\;\;\;\;\;\;\;\;\;\;\overline{\Lambda }%
^{1\;\dot{\alpha}}\;=\;Tr\left( \overline{c}\overline{F}^{\dot{\alpha}%
}\right) \;+\;Tr\left( c\overline{R}^{\dot{\alpha}}c\right) \;,
\label{lambda-sol}
\end{equation}
where $\overline{R}^{\dot{\alpha}}c$, $R^\alpha \overline{c}$ are given in
eqs.(\ref{R-action}).

In the same way we have 
\begin{eqnarray}
R^\alpha \;\omega _{\;\alpha }^2\; &=&R^\alpha \;\zeta _\alpha \;\omega
^3\;=\;\zeta _\alpha \;R^\alpha \;\omega ^3\;=\zeta _\alpha \;s\;\Lambda
^{1\;\alpha }\;  \nonumber \\
&=&\;s\;\left( \zeta _\alpha \;\Lambda ^{1\;\alpha }\right) \;+\;D_\alpha
\;\Lambda ^{1\;\alpha }\;,
\end{eqnarray}
showing that $R^\alpha \;\omega _{\;\alpha }^2$ is a trivial BRS cocycle
plus a total superspace derivative. The same conclusions hold for $\overline{%
R}_{\dot{\alpha}}\;\omega ^3$ and $\overline{R}_{\dot{\alpha}}\;\overline{%
\omega }^{2\;\dot{\alpha}}$ and can be extended by similar arguments to
include the $G$-terms $G_{\dot{\alpha}}^\alpha \;\omega ^3$ and $G_{\dot{%
\alpha}}^\alpha \;\widetilde{\omega }_{\;\;\alpha }^{2\;\dot{\alpha}}$.

The final result is that the equations (\ref{desc-count}) can be rewritten
without the explicit presence of the operators $R$ and $G$, yielding thus
the final version of the superspace descent equations for the invariant
action, \textit{i.e.} 
\begin{equation}
\begin{array}{l}
s\;\left( \omega ^0+\frac 14\overline{\zeta }_{\dot{\alpha}}\;\overline{%
\Lambda }^{1\;\dot{\alpha}}+\frac 14\zeta ^\alpha \;\Lambda _{\;\alpha
}^1\right) \;=-\frac 12\;D^\alpha \;\left( \omega _{\;\;\alpha
}^{1\;}\;+\;\frac 12\Lambda _\alpha ^1\right) \\ 
\;\;\;\;\;\;\;\;\;\;\;\;\;\;\;\;\;\;\;\;\;\;\;\;\;\;\;\;\;\;\;\;\;\;\;\;\;\;%
\;\;\;\;\;\;\;\;\;\;\;\;+\frac 12\;\overline{D}_{\dot{\alpha}}\;\left( 
\overline{\omega }^{1\;\dot{\alpha}}\;\;-\;\frac 12\overline{\Lambda }^{1\;%
\dot{\alpha}}\right) \;, \\ 
\\ 
s\;\left( \overline{\omega }_{\;\dot{\alpha}}^1\;\;-\;\frac 12\overline{%
\Lambda }_{\;\dot{\alpha}}^1\right) \;\;=-\frac 12\;\overline{D}_{\dot{\alpha%
}}D^\alpha \;\omega _{\;\alpha }^{2\;}\;\;-\;D^\alpha \overline{D}_{\dot{%
\alpha}}\;\omega _{\;\alpha }^{2\;}\;-\frac 12\;D^2\;\overline{\omega }_{\;\;%
\dot{\alpha}}^2\;, \\ 
\\ 
s\;\left( \omega ^{1\;\alpha }\;+\;\frac 12\Lambda ^{1\;\alpha }\right)
\;=\;\frac 12\;D^\alpha \overline{D}_{\dot{\alpha}}\;\overline{\omega }^{2\;%
\dot{\alpha}}\;\;+\;\overline{D}_{\dot{\alpha}}D^\alpha \;\overline{\omega }%
^{2\;\dot{\alpha}}+\;\frac 12\;\overline{D}^2\;\omega ^{2\;\alpha }\;, \\ 
\\ 
s\;\overline{\omega }_{\;\dot{\alpha}}^2\;\;=\;-\overline{D}_{\dot{\alpha}%
}\;\omega ^3\;, \\ 
\\ 
s\;\omega ^{2\;\alpha }\;=\;-D^\alpha \;\omega ^3\;, \\ 
\\ 
s\;\omega ^3\;\;\;=\;0\;.
\end{array}
\label{desc-count1}
\end{equation}

In particular the first equation of the above system explicitly shows that
the invariant action ${\cal{L}}^0$ can be identified with 
\begin{equation}
{\cal{L}}^0\;=\;\omega ^0+\frac 14\overline{\zeta }_{\dot{\alpha}}\;%
\overline{\Lambda }^{1\;\dot{\alpha}}+\frac 14\zeta ^\alpha \;\Lambda
_{\;\alpha }^1\;.  \label{YMaction}
\end{equation}
The above expression has to be understood modulo an exact BRS cocycle or a
total superspace derivative. Its nontriviality relies on the nontriviality
of the ghost cocycle (\ref{ghost-cocycle-c3}), as one can show by using a
well known standard cohomological argument\cite{Russ1,Russ2}. Recalling then
the expressions (\ref{sol-count}), (\ref{lambda-sol}), for ${\cal{L}}^0$
we get 
\[
{\cal{L}}^0\;=\;\frac 14Tr\;\left( \varphi ^\alpha \;F_\alpha \right)
\;+\;\frac 14Tr\;\left( \overline{\varphi }_{\dot{\alpha}}\;\overline{F}^{%
\dot{\alpha}}\right) \;, 
\]
which when integrated on the full superspace $d^4x\;d^2\theta \;d^2\overline{%
\theta }$ yields the familiar N=1 supersymmetric invariant Yang-Mills
lagrangian\footnote{%
We recall here the useful superspace identity $\int d^4x\;d^2\theta \;d^2%
\overline{\theta }\;=\;\int d^4x\;d^2\theta \;\overline{D}^2$.}: 
\[
S_{YM}=\int d^4x\;d^2\theta \;d^2\overline{\theta }\;{\cal{L}}^0=\frac
14\int d^4x\;d^2\theta \;Tr\;F^\alpha F_\alpha \;+\;\frac 14\int d^4x\;d^2%
\overline{\theta }\;Tr\;\overline{F}_{\dot{\alpha}}\overline{F}^{\dot{\alpha}%
}. 
\]

\section{Descent Equations for the U(1) Anomaly}

As already remarked in the Introduction the BRS consistency condition for
the chiral U(1) axial anomaly reads\cite{pig3,pigbook} 
\begin{equation}
s\;\int d^4x\;d^2\overline{\theta }\;K^0=\;0\;\Longrightarrow \;s\;K^0\;=\;%
\overline{D}_{\dot{\alpha}}\;\overline{K}^{1\dot{\alpha}}\;,
\label{WZ-chiral}
\end{equation}
where $K^0$ and $\;\overline{K}^{1\dot{\alpha}}$ have dimensions two and
three half and ghost numbers zero and one respectively. $K^0$ has thus the
same quantum numbers of the invariant action considered in the previous
section, the only difference lying in the fact that the superspace measure, 
\textit{i.e.}$\;d^4x\;d^2\overline{\theta }$, is now chiral instead of the
vector one $d^4x\;d^2\overline{\theta }\;d^2\theta $. Therefore the descent
equations for $K^0$ are obtained by performing the chiral limit of the
vector equations (\ref{desc-count}). Acting indeed with the BRS operator on
the second equation of the condition (\ref{WZ-chiral}), we obtain 
\begin{equation}
\overline{D}_{\dot{\alpha}}\;s\;\overline{K}^{1\dot{\alpha}}\;=\;0\;.
\label{step-WZchiral}
\end{equation}
Using then the results given in the App.A, it follows that the general
solution of the equation (\ref{step-WZchiral}) is given by 
\begin{eqnarray}
s\;\overline{K}^{1\dot{\alpha}}\; &=&\;\left( \overline{D}^{\dot{\alpha}%
}\;D^\alpha \;+\;2\;\;D^\alpha \;\overline{D}^{\dot{\alpha}}\right) K_\alpha
^2\;,  \label{dois-WZchiral} \\
\overline{D}^2\;K_\alpha ^2\; &=&\;0\;,  \nonumber
\end{eqnarray}
where $K_\alpha ^2$ is of dimension one half and ghost number two. Again,
acting with the BRS operator on the eq.(\ref{dois-WZchiral}) one gets 
\begin{equation}
\left( \overline{D}^{\dot{\alpha}}\;D^\alpha \;+\;2\;\;D^\alpha \;\overline{D%
}^{\dot{\alpha}}\right) \;s\;K_\alpha ^2\;=\;0\;,  \label{tres-WZchiral}
\end{equation}
which according to the App.A implies 
\begin{eqnarray*}
s\;\;K_\alpha ^2\; &=&\;D^\alpha \;K^3\;, \\
\overline{D}^2\;D^\alpha \;K^3\; &=&\;0\;,\;\;\;\;\;\;\;D^2\;\overline{D}^{%
\dot{\alpha}}\;K^3\;=\;0\;,
\end{eqnarray*}
with $K^3$ of dimension zero and ghost number three. Finally, from 
\[
D^\alpha \;s\;K^3\;=\;0\;, 
\]
it follows that 
\[
s\;K^3\;=\;0\;. 
\]
Summarizing, the superspace descent equations for the U(1) chiral axial
anomaly are 
\begin{equation}
\begin{array}{lll}
s\;K^0 & = & \overline{D}_{\dot{\alpha}}\overline{K}_{\;}^{1\;\dot{\alpha}%
}\;, \\ 
&  &  \\ 
s\;\overline{K}_{\;\dot{\alpha}}^1\; & = & \left( 2D^\alpha \overline{D}_{%
\dot{\alpha}}\;+\;\overline{D}_{\dot{\alpha}}D^\alpha \right) \;K_{\;\alpha
}^{2\;}\;, \\ 
&  &  \\ 
s\;K^{2\;\alpha } & = & D^\alpha \;K^3\;, \\ 
&  &  \\ 
s\;K^3\; & = & \;0\;,
\end{array}
\label{desc-chiral}
\end{equation}
with the constraints 
\begin{equation}
\begin{array}{c}
\overline{D}^2\;K_\alpha ^2\;=\;0\;, \\ 
\\ 
\overline{D}^2\;D^\alpha \;K^3\;=\;D^2\;\overline{D}^{\dot{\alpha}%
}\;K^3\;=\;0\;.
\end{array}
\label{constraints}
\end{equation}
Recalling then the result of the previous section, for $K^3$ we have 
\begin{equation}
K^3\;=\;\;\left( Tr\frac{c^3}3\;+\;Tr\frac{\overline{c}^3}3\right)
+\;s\;\Delta ^2\;,  \label{newc3}
\end{equation}
for some local power series $\Delta ^2$. It is interesting to observe that
in this case the constraints (\ref{constraints}) fix completely the trivial
part of $K^3$, giving for instance 
\[
s\;\Delta ^2\;=\;0\;. 
\]
Acting with the operator $\zeta _\alpha $ on both sides of the last of the
eqs.(\ref{desc-chiral}) and making use of the decomposition (\ref{super-dec}%
), for $K_\alpha ^2\;$ one gets 
\[
K_\alpha ^2\;=\;-\zeta _\alpha \;K^3\;+\;s\;\Delta _\alpha ^1\;. 
\]

Once more, it is not difficult to prove that the imposition of the
constraints (\ref{constraints}) yields a unique expression for $\Delta
_\alpha ^1$, \textit{i.e.} 
\[
\Delta _\alpha ^1\;=\;Tr\left( c\;\varphi _\alpha \right) \;, 
\]
so that for $K_\alpha ^2$ we get 
\[
\;K_\alpha ^2\;=\;\;Tr\left( c\;D_\alpha \;c\right) \;. 
\]
One sees thus that in the chiral case, due to the constraints (\ref
{constraints}), the trivial BRS contributions are uniquely fixed at the
lowest levels of the descent equations. Repeating now the same procedure and
making use of the relations (\ref{Gdef}) for $\overline{K}^{1\dot{\alpha}}$
one obtains 
\begin{equation}
\overline{K}^{1\dot{\alpha}}\;=\;G^{\alpha \dot{\alpha}}\;K_\alpha ^2\;-\;%
\overline{\Lambda }^{1\dot{\alpha}}\;+\;D^\alpha \overline{D}^{\dot{\alpha}%
}\;Tr\left( c\;\varphi _\alpha \right) \;+\;Tr\left( \overline{c}\;\overline{%
F}^{\dot{\alpha}}\right) \;\;+\;s\;\Delta ^{0\dot{\alpha}}\;,  \label{K1}
\end{equation}
where the cocycle $\overline{\Lambda }^{1\dot{\alpha}}\;$is the same as in
eq.(\ref{R-trivial}), \textit{i.e.} 
\[
\overline{\Lambda }^{1\dot{\alpha}}\;=\;Tr\left( c\;\overline{R}^{\dot{\alpha%
}}\;c\right) \;+\;Tr\left( \overline{c}\;\overline{F}^{\dot{\alpha}}\right)
\;. 
\]
It follows thus that 
\begin{equation}
\overline{K}^{1\dot{\alpha}}\;=-\;2\;Tr\left( D^\alpha \;c\;\overline{D}^{%
\dot{\alpha}}\;\varphi _\alpha \right) \;+\;s\;\Delta ^{0\dot{\alpha}}\;.
\label{newK1}
\end{equation}
Finally, acting with the operator $\overline{\zeta }_{\dot{\alpha}}$ on both
sides of the equation 
\[
\begin{array}{lll}
s\;\overline{K}_{\;\dot{\alpha}}^1\; & = & \left( 2D^\alpha \overline{D}_{%
\dot{\alpha}}\;+\;\overline{D}_{\dot{\alpha}}D^\alpha \right) \;K_{\;\alpha
}^{2\;}\;,
\end{array}
\]
for the last level $K^0$ we find 
\[
K^0\;=\;-\;\overline{\zeta }_{\dot{\alpha}}\;\overline{K}^{1\dot{\alpha}%
}\;+\;Tr\left( 2\varphi ^\alpha \;F_\alpha \;+\;\overline{D}_{\dot{\alpha}%
}\;\varphi ^\alpha \;\overline{D}^{\dot{\alpha}}\;\varphi _\alpha \right)
\;, 
\]
reproducing the well known expression for the U(1) supersymmetric chiral
anomaly 
\[
K^0\;=\;\;Tr\left( 2\varphi ^\alpha \;F_\alpha \;-\;\overline{D}_{\dot{\alpha%
}}\;\varphi ^\alpha \;\overline{D}^{\dot{\alpha}}\;\varphi _\alpha \right)
\;-\;\overline{D}_{\dot{\alpha}}\;\Delta ^{0\dot{\alpha}}\;. 
\]
Let us conclude by remarking that the expressions of the cocycles $K^3$, $%
K_\alpha ^2$, $\overline{K}^{1\dot{\alpha}}$ and $K^0$ found here are
completely equivalent to those of \cite{pig3}, \textit{i.e.} the difference
is an exact BRS cocycle or a total superspace derivative.

\section{The Supersymmetric Gauge Anomaly}

As the last example of our superspace analysis let us consider the case of
the supersymmetric gauge anomaly. As usual let us first focus on the
derivation of the corresponding descent equations. The latters, as mentioned
in the Introduction and in the Section IV , can be obtained by adding to the
right hand side of the generalized equation (\ref{desc2}) an appropriate
extra term. The presence of this term actually stems from the BRS triviality%
\cite{pig2} of the pure ghost cocycles $\left( Tr\;c^{2n+1}-Tr\;\overline{c}%
^{2n+1}\right) $, $n\geq 1$,$\;$ 
\begin{equation}
s\;\Omega ^{2n}=\;Tr\;\frac{c^{2n+1}}{2n+1}\;-Tr\;\frac{\overline{c}^{2n+1}}{%
2n+1}\;,  \label{triviality}
\end{equation}
$\Omega \;^{2n}$ being a local dimensionless functional of $(\phi ,c,%
\overline{c})$ with ghost number $2n$. Acting in fact with the operator $%
e^\delta $ on both sides of eq.(\ref{triviality}) and recalling the
definitions (\ref{Atilde-con}) and (\ref{Abartilde-con}) we get the desired
modified version of the generalized superspace equation (\ref{desc2}) we are
looking for, 
\begin{eqnarray}
\widetilde{d}\;\widetilde{\Omega }\; &=&Tr\;\frac{\widetilde{A}^{2n+1}}{2n+1}%
\;-Tr\;\frac{\widetilde{\overline{A}}^{2n+1}}{2n+1}\;\;,  \label{dOmega} \\
\widetilde{\Omega }\;\; &=&\;e^\delta \;\Omega ^{2n}\;.  \nonumber
\end{eqnarray}

The descent equations for the gauge anomaly follows then from eq.(\ref
{dOmega}) when $n=2$, \textit{i.e.} 
\[
\widetilde{d}\;\widetilde{\Omega }=\;\frac 15\;Tr\left( \widetilde{A}^5\;-\;%
\widetilde{\overline{A}}^5\right) \;, 
\]
\begin{equation}
\widetilde{\Omega }=\;e^\delta \;\Omega ^4\;.  \label{Atilde}
\end{equation}
To see that the above equation characterizes indeed the gauge anomaly let us
write it in components. Expanding $\widetilde{\Omega }$ in the global
parameters $(e^\alpha ,\overline{e}^{\dot{\alpha}},\widetilde{e}_\alpha ^{%
\dot{\alpha}})$ 
\begin{equation}
\begin{array}{ccc}
\widetilde{\Omega } & = & \;\Omega ^4\;+\;\;\Omega ^{3\;\alpha }\;e_\alpha
\;+\;\;\overline{\Omega }_{\;\dot{\alpha}}^3\;\overline{e}^{\dot{\alpha}%
}\;+\;\widetilde{\Omega }_{\;\;\dot{\alpha}}^{3\;\alpha }\;\widetilde{e}%
_\alpha ^{\dot{\alpha}}\;+\;\widetilde{\Omega }_{\;\;\dot{\alpha}%
}^{2\;\alpha }\;e_\alpha \;\overline{e}^{\dot{\alpha}} \\ 
&  &  \\ 
&  & +\;\Omega ^{2\;\alpha }\;\widetilde{e}_{\alpha \dot{\alpha}}\;\overline{%
e}^{\dot{\alpha}}\;+\;\overline{\Omega }_{\;\dot{\alpha}}^2\;e^\alpha \;%
\widetilde{e}_\alpha ^{\dot{\alpha}}\;+\;\Omega ^1\;e^\alpha \;\widetilde{e}%
_{\alpha \dot{\alpha}}\;\overline{e}^{\dot{\alpha}}\;,
\end{array}
\label{Otilde}
\end{equation}
and eliminating the $G$ and $R$ terms as done in Subsect. IV.1 we get the
known descent equations for the superspace gauge anomaly\cite{pig2,porrati} 
\begin{equation}
\begin{array}{lll}
s\;\;\Omega ^1 & = & D^\alpha \;\Omega _\alpha ^2\;+\;\overline{D}_{\dot{%
\alpha}}\;\Omega ^{2\dot{\alpha}}\;, \\ 
&  &  \\ 
s\;\;\Omega _\alpha ^2 & = & -\overline{D}^2\Omega _\alpha ^3\;+\;\left( 2%
\overline{D}_{\dot{\alpha}}D_\alpha \;+\;D_\alpha \overline{D}_{\dot{\alpha}%
}\right) \;\overline{\Omega }^{3\dot{\alpha}}\; \\ 
&  &  \\ 
&  & \;\;\;\;\;\;\;+\;2\;Tr\left( \left( D_\alpha \overline{D}_{\dot{\alpha}}%
\overline{c}\right) \left( \overline{c}\overline{D}^{\dot{\alpha}}\overline{c%
}\;+\;\overline{D}^{\dot{\alpha}}\overline{c}\;\overline{c}\right) \right)
\;, \\ 
&  &  \\ 
s\;\overline{\Omega }^{2\dot{\alpha}} & = & D^2\overline{\Omega }^{3\dot{%
\alpha}}\;-\;\left( 2D^\alpha \overline{D}_{\dot{\alpha}}\;+\;\overline{D}_{%
\dot{\alpha}}D^\alpha \right) \;\Omega _\alpha ^3\;\; \\ 
&  &  \\ 
&  & \;\;\;\;\;\;\;-\;2\;Tr\left( \left( \overline{D}^{\dot{\alpha}}D^\alpha
c\right) \left( cD_\alpha c\;+\;D_\alpha c\;c\right) \right) \;, \\ 
&  &  \\ 
s\;\Omega _\alpha ^3\; & = & D_\alpha \;\Omega ^4\;+\;Tr\left( c^3D_\alpha
c\right) \;, \\ 
&  &  \\ 
s\;\overline{\Omega }^{3\dot{\alpha}} & = & -\overline{D}^{\dot{\alpha}%
}\;\Omega ^4\;+\;Tr\left( \overline{c}^3\overline{D}^{\dot{\alpha}}\overline{%
c}\right) \;, \\ 
&  &  \\ 
s\;\Omega ^4 & = & \frac 15Tr\left( c^5\;-\;\overline{c}^5\right) \;.
\end{array}
\label{desc-gauge}
\end{equation}
One sees in particular that integrating the first equation of (\ref
{desc-gauge}) on superspace, the cocycle $\Omega ^1$ obeys exactly the BRS
consistency condition corresponding to the possible gauge breakings 
\[
s\;\int d^4x\;d^2\theta \;d^2\overline{\theta }\;\Omega ^1=\;0\;, 
\]
identifying therefore $\Omega ^1$ with the supersymmetric Yang-Mills anomaly.

In order to find a solution of the descent equations (\ref{desc-gauge}) we
use the same climbing procedure of the previous examples obtaining the
following nontrivial expressions 
\begin{equation}
\begin{array}{lll}
\Omega _\alpha ^3\; & = & -\;\zeta ^\alpha \;\Omega ^4\;-\;Tr\left( \varphi
^\alpha \;c^3\right) \;, \\ 
&  &  \\ 
\overline{\Omega }^{3\dot{\alpha}} & = & \overline{\zeta }_{\dot{\alpha}%
}\;\Omega ^4\;-\;Tr\left( \overline{\varphi }_{\dot{\alpha}}\;\overline{c}%
^3\right) \;, \\ 
&  &  \\ 
\Omega _\alpha ^2 & = & G_{\alpha \dot{\alpha}}\;\overline{\zeta }^{\dot{%
\alpha}}\;\Omega ^4\;+\;\;\overline{D}_{\dot{\alpha}}\;\overline{\zeta }^{%
\dot{\alpha}}\;\zeta _\alpha \;\Omega ^4 \\ 
&  &  \\ 
&  & \;\;-\;Tr\left( \overline{\varphi }_{\dot{\alpha}}\;\left( D_\alpha 
\overline{\varphi }^{\dot{\alpha}}\right) \overline{c}^2\;-\;\overline{%
\varphi }_{\dot{\alpha}}\;\overline{c}\left( D_\alpha \overline{\varphi }^{%
\dot{\alpha}}\right) \overline{c}\;+\;\overline{\varphi }_{\dot{\alpha}}\;%
\overline{c}^2\;D_\alpha \overline{\varphi }^{\dot{\alpha}}\right) \\ 
&  &  \\ 
&  & \;\;+2\;Tr\left( \left( D_\alpha \overline{\varphi }_{\dot{\alpha}%
}\right) \left( \overline{c}\overline{D}^{\dot{\alpha}}\overline{c}\;+\;%
\overline{D}^{\dot{\alpha}}\overline{c}\;\overline{c}\right) \right) \;, \\ 
&  &  \\ 
\overline{\Omega }^{2\dot{\alpha}} & = & G^{\alpha \dot{\alpha}}\;\zeta
_\alpha \;\Omega ^4\;\;+\;D^\alpha \;\overline{\zeta }^{\dot{\alpha}}\;\zeta
_\alpha \;\Omega ^4 \\ 
&  &  \\ 
&  & \;+\;Tr\left( \varphi ^\alpha \;\left( \overline{D}^{\dot{\alpha}%
}\varphi _\alpha \right) \overline{c}^2\;-\;\varphi ^\alpha \;\overline{c}%
\left( \overline{D}^{\dot{\alpha}}\varphi _\alpha \right) \overline{c}%
\;+\;\varphi ^\alpha \;\overline{c}^2\;\overline{D}^{\dot{\alpha}}\varphi
_\alpha \right) \\ 
&  &  \\ 
&  & \;-2\;Tr\left( \left( \overline{D}^{\dot{\alpha}}\varphi _\alpha
\right) \left( cD_\alpha c\;+\;D_\alpha c\;c\right) \right) \;,
\end{array}
\label{A2}
\end{equation}
and for the gauge anomaly 
\begin{equation}
\begin{array}{lll}
\Omega ^1 & = & 2\;\zeta ^\alpha \;G_{\alpha \dot{\alpha}}\;\overline{\zeta }%
^{\dot{\alpha}}\;\Omega ^4\; \\ 
&  &  \\ 
&  & +2\;Tr\left( F^\alpha \;c\;\varphi _\alpha \;-\;F^\alpha \;\varphi
_\alpha \;c\;+\;\left( \overline{D}_{\dot{\alpha}}\;\varphi ^\alpha \right)
\left( \overline{D}^{\dot{\alpha}}\;\varphi _\alpha \right) c\right) \\ 
&  &  \\ 
&  & -2\;Tr\left( \overline{F}_{\dot{\alpha}}\;\overline{c}\;\overline{%
\varphi }^{\dot{\alpha}}\;-\;\overline{F}_{\dot{\alpha}}\;\overline{\varphi }%
^{\dot{\alpha}}\;\overline{c}\;+\;\left( D^a\;\overline{\varphi }_{\dot{%
\alpha}}\right) \left( D_\alpha \;\overline{\varphi }^{\dot{\alpha}}\right) 
\overline{c}\right) \;.
\end{array}
\label{A0}
\end{equation}

One should observe that the explicit final expression for the gauge anomaly
depends on the knwoledge of the cocycle $\Omega ^4$ solution of the last of
the descent equations (\ref{desc-gauge}). This point is particularly
important and deserves some further clarifying remarks.

\subsection{Nonpolynomial Character of The Gauge Anomaly}

It is known that due to a theorem by Ferrara, Girardello, Piguet and Stora 
\cite{fgps}, the superspace gauge anomaly cannot be expressed as a
polynomial in the variables $(\varphi _\alpha ,\lambda _\alpha \equiv
e^\varphi D_\alpha \;e^{-\varphi })$ and their covariant derivatives. In
fact all the known superspace closed expressions of the gauge anomaly so far
obtained by means of homotopic transgression procedures\cite{niels,ggs,bpt,gm}
show up an highly nonpolynomial character in the gauge superconnetion. On
the other hand in our approach the simple knowledge of the cocycle $\Omega
^4 $ would produce a closed expression for the supersymmetric gauge anomaly
without any homotopic integral. Of course this would imply a deeper
understanding of this anomaly. It is not difficult however to convince
oneself that solving the equation 
\begin{equation}
\begin{array}{lll}
s\;\Omega ^4 & = & \frac 15Tr\left( c^5\;-\;\overline{c}^5\right) \;
\end{array}
\label{last point}
\end{equation}
is not an easy task. This is actually due to the BRS transformation of the
vector superfield $\phi $%
\[
s\;e^\phi \;=\;e^\phi \;c\;-\;\overline{c}\;e^\phi \;. 
\]
which when written in terms of $\phi $ takes the highly complex form\cite
{pigbook} 
\begin{equation}
s\;\phi \;=\;\frac 12{\cal{L}}_\phi \;(c+\overline{c})\;+\;\frac 12%
{\cal{L}}_\phi \;\left[ \coth \;\left( \frac{{\cal{L}}_\phi }2\right)
\right] \;(c-\overline{c})\;,  \label{phi-transf}
\end{equation}
where 
\[
{\cal{L}}_\phi \;\cdot \;=\;[\phi ,\;\cdot \;]\;, 
\]
and 
\[
\coth \;\left( \frac{{\cal{L}}_\phi }2\right) \;=\;\frac{e^{\frac{{\cal{L}}_\phi }2}\;+\;e^{-\frac{{\cal{L}}_\phi }2}}{e^{\frac{{\cal{L}}_\phi }%
2}\;-\;e^{-\frac{{\cal{L}}_\phi }2}}\;. 
\]
The formula (\ref{phi-transf}) can be expanded in powers of $\phi $,
allowing to solve the equation (\ref{last point}) order by order in the
vector superfield $\phi $. For instance, in the first approximation which
corresponds to the abelian limit of retaining only the linear terms of the
BRS transformations, \textit{i.e.} 
\[
s\;\rightarrow \;s_{ab} 
\]
with 
\begin{eqnarray*}
s_{ab}\;\phi \; &=&\;c\;-\;\overline{c}\;, \\
s_{ab}\;c\; &=&\;s_{ab}\;\overline{c}\;=\;0\;,
\end{eqnarray*}
one easily checks that 
\begin{equation}
Tr\left( c^5\;-\;\overline{c}^5\right) \;=\;s_{ab}\;Tr\left( \phi \left(
c^4\;+\;c^3\;\overline{c}\;+\;c^2\;\overline{c}^2\;+\;c\;\overline{c}^3\;+\;%
\overline{c}^4\right) \right) \;,  \label{abelian-approx}
\end{equation}
which shows indeed the BRS triviality\cite{book} of $Tr\left( c^5\;-\;%
\overline{c}^5\right) $.

Up to our knowledge a closed exact form for $\Omega ^4$ has not yet been
established. In other words, due to the theorem of Ferrara, Girardello,
Piguet and Stora\cite{fgps}, the nonpolynomiality of the supersymmetric
gauge anomaly directly relies on the nonpolynomial nature of the cocycle $%
\Omega ^4$. Any progress in this direction will be reported as soon as
possible.

Let us conclude this section by giving the explicit expression of the gauge
anomaly (\ref{A0}) up to the second order in the vector field $\phi $, 
\textit{i.e.} 
\begin{equation}
\begin{array}{lll}
\Omega ^1 & = & -2\;Tr\left( D^\alpha \phi \;\overline{D}^2D_\alpha \phi
\;c\;+\;\overline{D}^2D^\alpha \phi \;D_\alpha \phi \;c\;-\;\left( \overline{%
D}_{\dot{\alpha}}D^\alpha \phi \right) \left( \overline{D}^{\dot{\alpha}%
}D_\alpha \phi \right) c\right) \\ 
&  &  \\ 
&  & +2\;Tr\left( \overline{D}_{\dot{\alpha}}\phi \;D^2\overline{D}^{\dot{%
\alpha}}\phi \;\overline{c}\;+\;D^2\overline{D}_{\dot{\alpha}}\phi \;%
\overline{D}^{\dot{\alpha}}\phi \;\overline{c}\;-\;\left( D^a\overline{D}_{%
\dot{\alpha}}\phi \right) \left( D_\alpha \overline{D}^{\dot{\alpha}}\phi
\right) \overline{c}\right) \;,
\end{array}
\label{A1}
\end{equation}
which is easily recognized to be equivalent to that of ref.\cite{pig2}. One
should also observe that the above expression do not receive contributions
from the term $\Omega ^4$ since they are at least of the order three in $%
\phi $, as it can be checked by applying the combination $\zeta ^\alpha
\;G_{\alpha \dot{\alpha}}\;\overline{\zeta }^{\dot{\alpha}}$ on the cocycle
of the eq.(\ref{abelian-approx}).

\section{Conclusion}

The supersymmetric version of the descent equations for the four dimensional
N=1 Super-Yang-Mills gauge theories can be analysed by means of the
introduction of two operators $\zeta ^\alpha $ and $\overline{\zeta }^{\dot{%
\alpha}}$which decompose the supersymmetric derivatives $D^\alpha \;$and $%
\overline{D}^{\dot{\alpha}}$ as BRS commutators. These operators provide an
algebraic setup for a systematic derivation of the superspace descent
equations. In addition they allow to cast both the supersymmetric BRS
transformations and the descent equations into a very suggestive zero
curvature formalism in superspace.

\section*{Acknowledgements}

We are grateful to Olivier Piguet for many valuable discussions and private
communications. The Conselho Nacional de Desenvolvimento Cient\'\i fico e
Tecnol\'ogico CNPq-Brazil and SR2-UERJ are gratefully acknowledged for the
financial support.

\newpage\ 

\appendix 

\section{Appendix}

We list here the superspace algebraic solutions \cite{pig2,pig3,pigbook,pc}
of some equations needed for the analysis of the supersymmetric descent
equations. All these solutions are built up by superfields. They have always
to be understood modulo terms which automatically solve the corresponding
equations but cannot be written in the same algebraic form as the
solutions.The existence of such particular terms strongly depends on the
superfield content of the particular model under consideration.

The first result states that the solution of the superspace equation 
\[
\overline{D}^2\;Q\;=\;0\;, 
\]
can be generically written as 
\[
Q\;=\overline{D}_{\dot{\alpha}}\;{\mathcal{M}}^{\dot{\alpha}}\;, 
\]
for some superfield ${\cal{M}}^{\dot{\alpha}}\;$.

The second important result concerns the solution of the following equation 
\[
\left( 2\overline{D}_{\dot{\alpha}}D_\alpha \;+\;D_\alpha \overline{D}_{\dot{%
\alpha}}\right) \;\overline{Q}^{\dot{\alpha}}\;=\;\overline{D}^2\;Q_\alpha
\;. 
\]
For the superfields $\overline{Q}^{\dot{\alpha}}$ and $Q_\alpha \;$we have
now 
\[
\begin{array}{ccc}
\overline{Q}^{\dot{\alpha}} & = & \overline{D}^{\dot{\alpha}}\;\mathcal{M}\;,
\\ 
&  &  \\ 
Q_\alpha & = & -D_\alpha \;\mathcal{M}\;,
\end{array}
\]
with $\mathcal{\;M}$ an arbitrary superfield. Let us observe that in this
case the term $Tr\left( cD_\alpha c\right) $, due to the fact that the ghost 
$c$ is a chiral superfield, is automatically annihilated by the operator $%
\overline{D}^2$. Therefore it must be included in the expression given for $%
Q_\alpha \;$although it cannot be written as a total superspace derivative.

Considering now the equation 
\[
D^\alpha \;Q_\alpha \;=\;\overline{D}_{\dot{\alpha}}\;\overline{Q}^{\dot{%
\alpha}}\;, 
\]
we have 
\begin{equation}
\begin{array}{ccc}
Q_\alpha \; & = & -\overline{D}^2\;{\mathcal{P}}_\alpha \;+\;\left( 2\overline{%
D}_{\dot{\alpha}}D_\alpha \;+\;D_\alpha \overline{D}_{\dot{\alpha}}\right) \;%
\overline{\mathcal{P}}^{\dot{\alpha}}\;+\;D^\beta \;{\mathcal{N}}_{\left(
\alpha \beta \right) }\;, \\ 
&  &  \\ 
\overline{Q}^{\dot{\alpha}} & = & -D^2\;\overline{\mathcal{P}}^{\dot{\alpha}%
}\;+\;\left( 2D^\alpha \overline{D}^{\dot{\alpha}}\;+\;\overline{D}^{\dot{%
\alpha}}D^\alpha \right) \;{\mathcal{P}}_\alpha \;+\;\overline{D}_{\dot{\beta}}%
\overline{\mathcal{N}}^{\left( \dot{\alpha}\dot{\beta}\right) }\;,
\end{array}
\label{algebric}
\end{equation}
with $\mathcal{P}_\alpha \;$ and $\mathcal{N}_{\left( \alpha \beta \right) }$
appropriate superfields. Of course the existence of the symmetric superfield 
$\mathcal{N}_{\left( \alpha \beta \right) }$ depends on the dimension and on
the ghost number of $Q_\alpha $. For instance in the case of the vector
descent equations (\ref{desc-count1}) in which $Q_\alpha $ corresponds to $%
s(\omega _\alpha ^1+\frac 12\Lambda _\alpha ^1)$, it is not difficult to
check that $\mathcal{N}_{\left( \alpha \beta \right) }$ is automatically
absent due to the quantum numbers of the problem.

In particular, in the case of the chiral descent equations considered in
Section V, eq.(\ref{algebric}) imply that the most general solution of the eq.(%
\ref{step-WZchiral}) is given indeed by 
\[
s\;\overline{K}^{1\dot{\alpha}}\;=\;\left( \overline{D}^{\dot{\alpha}%
}\;D^\alpha \;+\;2\;\;D^\alpha \;\overline{D}^{\dot{\alpha}}\right) K_\alpha
^2\;, 
\]
with the constraint 
\[
\overline{D}^2\;K_\alpha ^2\;=\;0\;. 
\]

\section{Appendix}

In this appendix we summarize some useful results concerning the BRS
superspace cohomology for the N=1 supersymmetric Yang-Mills gauge theories.
The various BRS cohomolgy classes are labelled by the ghost number $g$ and
by the spinor indices.

The following results hold\cite{pig2,pig3,porrati}:

\begin{enumerate}
\item  The BRS cohomology is empty in the space of the invariant local power
series $A^g$ with dimension $2$ and positive ghost number $g$.

\item  The cohomology classes corresponding to local BRS invariant cocycles $%
A_\alpha ^g$ or $\overline{A}_{\dot{\alpha}}^g$ with dimension $\frac 32$
and ghost number $g$ = 1, 2 or 3 are empty.

\item  The cohomology classes in the space of the BRS invariant local power
series $A_\alpha ^g$ or $\overline{A}_{\dot{\alpha}}^g$ with dimension $%
\frac 12$ and ghost number $g$ greater than zero are empty.

\item  The BRS cohomology classes in the space of the local power series $%
A^g $ with dimension $0$, ghost number $g$ and at least of order $g+1$ in
the fields are empty.

\item  Any invariant object $A^g$ with dimension $0$ and even ghost number $%
g $ greater than zero and of order $g$ in the fields is BRS trivial.
\end{enumerate}

In particular it turns out that in the pure ghost sector the BRS cohomology
classes are given by polynomials built up with monomials of the type

\begin{equation}
Tr\frac{c^{2n+1}}{2n+1}\;,\;\;\;\;\;\;\;n\geq 1\;,  \label{c-cocycles}
\end{equation}
or

\begin{equation}
Tr\frac{\overline{c}^{2n+1}}{2n+1}\;,\;\;\;\;\;\;\;n\geq 1\;.
\label{bc-cocycles}
\end{equation}
We remark also that the two expressions above (\ref{c-cocycles}) and (\ref
{bc-cocycles}) do not actually define different cohomology classes. Instead
they are equivalent, due to the triviality\cite{pig2,pigbook} of the
combination

\[
Tr\frac{c^{2n+1}}{2n+1}\;-\;Tr\frac{\overline{c}^{2n+1}}{2n+1}\;=\;s\;\Omega
^{2n}\;, 
\]
for some local power series $\Omega ^{2n}$. This result implies that the
expression (\ref{c-cocycles}) and (\ref{bc-cocycles}) are related each other
by means of an exact BRS\ cocycle.

\section{Appendix}

In this appendix we show that the off-shell closure of the algebra (\ref
{free-algebra}) can be recovered in a simple way by introducing an
appropriate external field $\eta $. Let indeed be $\eta $ a superfield with
dimension 2 and ghost number -1, whose BRS transformation reads 
\begin{eqnarray*}
s\;\eta \; &=&\;\left[ \eta \;,\;c\right] \;+2\;\left( D\;F\;+\;\left[
\varphi \;,\;F\right] \right) \;\;, \\
s^2\;\eta \; &=&\;0\;.
\end{eqnarray*}
Modifying now the operator $\zeta $ in such a way\ that 
\[
\zeta \;F\;=\;-\frac 12\;\eta \;, 
\]
it is easily verified that the commutator (\ref{on-shell}) 
\[
\left[ \zeta \;,\;s\right] \;F\;=\;-\zeta \;\left[ c\;,\;F\right] \;+\frac
12\;s\;\eta \;=D\;F\;, 
\]
gives now the covariant derivative of $F$ without making use of the
equations of motion, closing therefore the algebra (\ref{free-algebra})
off-shell. Let us conclude by also remarking that the external field $\eta $
cannot contribute to the BRS cohomology classes relevant for the examples
considered in the previous Sections due to its ghost number and to its
dimension.

\section{Appendix}

This appendix is devoted to some technical details concerning the introduction of 
the global parameters $e^\alpha $, $\overline{e}^{\dot{\alpha}}$ and $%
\widetilde{e}^{\alpha \dot{\alpha}}$. As it has been already underlined, these 
parameters have been introduced in order to project the algebraic relations 
(\ref{delta definition})-(\ref{commutators}) on the equations of motion of 
N=1 super Yang-Mills. This has been achieved by requiring that the conditions 
(\ref{basis}) are fulfilled. 
In addition, as it is apparent from the construction of Section IV, the introduction
of these global parameters, while collecting all the algebraic
relations (\ref{delta definition})-(\ref{commutators}) into a unique
extended generalized operator $\widetilde{d}$ (see equation (\ref{dtilde1}%
)), is of great relevance in order to cast the BRS transformations of the
superfields and the full system of superspace consistency conditions (\ref
{desc-count1}), (\ref{desc-chiral}) and (\ref{desc-gauge}) into a unique
equation. The latter  has the meaning of a zero-curvature condition (see eqs.(\ref
{zero-curv})-(\ref{desc1})). 
Therefore, the expansion of the equation  
\[
\widetilde{d}\;\widetilde{\omega }\;=\;0\;, 
\]
in powers of the global parameters ($e^\alpha $, $\overline{e}^{\dot{\alpha}%
} $, $\widetilde{e}^{\alpha \dot{\alpha}}$) will automatically provide the
full set of the superspace consistency conditions for the cocycle $%
\widetilde{\omega }$. In this sense, the parameters ($e^\alpha $, $\overline{%
e}^{\dot{\alpha}}$, $\widetilde{e}^{\alpha \dot{\alpha}}$) can be iseen as a
suitable basis in superspace for the zero curvature formulation of $N=1$
supersymmetry. 

It is also worth remarking that this construction does not have the meaning
of collecting into an extended BRS operator the various classical symmetries
of a given gauge-fixed action, as done for instance in the case of the extended (%
$N\geq 2$) supersymmetric Yang-Mills theories \cite{W,M} \footnote{%
It is indeed rather simple to convince oneself that the operators ($\zeta
_\alpha $,$\overline{\zeta }_{\dot{\alpha}}$) do not actually represent
invariances of the gauge-fixed $N=1$ super Yang-Mills action.}. Rather, it
is closer to the zero-curvature formulation of the topological field
theories discussed by \cite{PR,zero-curv} in terms of the so-called
universal bundle.

\end{document}